\documentclass[twocolumn]{aastex631}

\usepackage{txfonts}

\newcommand{\target}{Gliese\,12\xspace}
\newcommand{\planet}{\target~b\xspace}

\newcommand{\mear}{M_{\oplus}}

\newcommand{\rear}{R_{\oplus}}

\usepackage{graphicx} 
\usepackage{natbib} 
\usepackage{enumerate}
\usepackage{xspace}
\usepackage{xcolor}
\usepackage{comment}

 \shorttitle{Transiting Earth-Sized Temperate Planet around \target}
\shortauthors{Kuzuhara, Fukui et al.}
\graphicspath{{./}{figures/}}

\begin{document}

\title{Gliese 12 b: A temperate Earth-sized planet at 12\,pc ideal for atmospheric transmission spectroscopy}

\correspondingauthor{Masayuki Kuzuhara}
\email{m.kuzuhara@nao.ac.jp}

\author[0000-0002-4677-9182]{Masayuki Kuzuhara}
\altaffiliation{These two authors contributed equally to this work.}
\affiliation{Astrobiology Center, NINS, 2-21-1 Osawa, Mitaka, Tokyo 181-8588, Japan}
\affiliation{National Astronomical Observatory of Japan, NINS, 2-21-1 Osawa, Mitaka, Tokyo 181-8588, Japan}

\author[0000-0002-4909-5763]{Akihiko Fukui}
\altaffiliation{These two authors contributed equally to this work.}
\affiliation{Komaba Institute for Science, The University of Tokyo, 3-8-1 Komaba, Meguro, Tokyo 153-8902, Japan}
\affiliation{Instituto de Astrof\'isica de Canarias (IAC), E-38200 La Laguna, Tenerife, Spain\label{IAC}}

\author[0000-0002-4881-3620]{John H. Livingston}
\affiliation{Astrobiology Center, NINS, 2-21-1 Osawa, Mitaka, Tokyo 181-8588, Japan}
\affiliation{National Astronomical Observatory of Japan, NINS, 2-21-1 Osawa, Mitaka, Tokyo 181-8588, Japan}
\affiliation{Astronomical Science Program, Graduate University for Advanced Studies (SOKENDAI), 2-21-1, Osawa, Mitaka, Tokyo, 181-8588, Japan}

\author[0000-0002-7349-1387]{Jos\'e A. Caballero}
\affiliation{Centro de Astrobiología, CSIC-INTA, Camino Bajo del Castillo s/n, E-28692 Villanueva de la Cañada, Madrid, Spain}

\author[0000-0002-6424-3410]{Jerome P. de Leon}
\affiliation{Department of Multi-Disciplinary Sciences, Graduate School of Arts and Sciences, The University of Tokyo, 3-8-1 Komaba, Meguro, Tokyo 153-8902, Japan}

\author[0000-0003-3618-7535]{Teruyuki Hirano}
\affiliation{Astrobiology Center, NINS, 2-21-1 Osawa, Mitaka, Tokyo 181-8588, Japan}
\affiliation{National Astronomical Observatory of Japan, NINS, 2-21-1 Osawa, Mitaka, Tokyo 181-8588, Japan}
\affiliation{Astronomical Science Program, Graduate University for Advanced Studies (SOKENDAI), 2-21-1, Osawa, Mitaka, Tokyo, 181-8588, Japan}

\author[0000-0002-8607-358X]{Yui Kasagi}
\affiliation{Astronomical Science Program, Graduate University for Advanced Studies (SOKENDAI), 2-21-1, Osawa, Mitaka, Tokyo, 181-8588, Japan}
\affiliation{Astrobiology Center, NINS, 2-21-1 Osawa, Mitaka, Tokyo 181-8588, Japan}
\affiliation{National Astronomical Observatory of Japan, NINS, 2-21-1 Osawa, Mitaka, Tokyo 181-8588, Japan}

\author[0000-0001-9087-1245]{Felipe Murgas}
\affiliation{Instituto de Astrof\'isica de Canarias (IAC), E-38200 La Laguna, Tenerife, Spain\label{IAC}}
\affiliation{Departamento de Astrof\'isica, Universidad de La Laguna (ULL), E-38206 La Laguna, Tenerife, Spain\label{ull}}

\author[0000-0001-8511-2981]{Norio Narita}
\affiliation{Komaba Institute for Science, The University of Tokyo, 3-8-1 Komaba, Meguro, Tokyo 153-8902, Japan}
\affiliation{Astrobiology Center, NINS, 2-21-1 Osawa, Mitaka, Tokyo 181-8588, Japan}
\affiliation{Instituto de Astrof\'isica de Canarias (IAC), E-38200 La Laguna, Tenerife, Spain\label{IAC}}

\author[0000-0002-5051-6027]{Masashi Omiya}
\affiliation{Astrobiology Center, NINS, 2-21-1 Osawa, Mitaka, Tokyo 181-8588, Japan}
\affiliation{National Astronomical Observatory of Japan, NINS, 2-21-1 Osawa, Mitaka, Tokyo 181-8588, Japan}

% ...

\begin{abstract}
Recent discoveries of Earth-sized planets transiting nearby M dwarfs have made it possible to characterize the atmospheres of terrestrial planets via follow-up spectroscopic observations. However, the number of such planets receiving low insolation is still small, limiting our ability to understand the diversity of the atmospheric composition and climates of temperate terrestrial planets.
We report the discovery of an Earth-sized planet transiting the nearby (12\,pc) inactive M3.0 dwarf \target (TOI-6251) with an orbital period ($P_{\rm{orb}}$) of 12.76\,days.
The planet, \planet, was initially identified as a candidate with an ambiguous $P_{\rm{orb}}$ from TESS data. 
We confirmed the transit signal and $P_{\rm{orb}}$ using ground-based photometry with MuSCAT2 and MuSCAT3, and validated the planetary nature of the signal using high-resolution images from Gemini/NIRI and Keck/NIRC2 as well as radial velocity (RV) measurements from the InfraRed Doppler
instrument on the Subaru 8.2\,m telescope and from CARMENES on the CAHA 3.5\,m telescope. 
X-ray observations with XMM-Newton showed the host star is inactive, with an X-ray-to-bolometric luminosity ratio of $\log L_{\rm X}/L_{\rm bol} \approx -5.7$.
Joint analysis of the light curves and RV measurements revealed that \planet has a radius of $0.96 \pm 0.05 \, R_\oplus$, a 3$\sigma$ mass upper limit of $3.9 \, M_\oplus$, and an equilibrium temperature of $315 \pm 6$\,K assuming zero albedo. 
The transmission spectroscopy metric (TSM) value of \planet is close to the TSM values of the TRAPPIST-1 planets, adding \planet to the small list of potentially terrestrial, temperate planets amenable to atmospheric characterization with JWST. 
\end{abstract}


\section{Introduction} \label{sec:intro}

M dwarfs are promising targets to explore terrestrial exoplanets because of their small masses and radii, which result in large reflex motions and deep transits even for small planets. In addition, the low bolometric luminosity of M dwarfs makes the habitable zone (HZ) closer to the host stars, making planets inside the HZ easier to observe compared to those inside the HZ around earlier-type stars.
These advantages motivated several ground-based exoplanet-hunting projects to target nearby M dwarfs using Doppler velocimeters such as the High-Accuracy Radial-velocity Planet Searcher \citep[HARPS;][]{2013A&A...556A.110B}, the InfraRed Doppler instrument \citep[IRD;][]{Harakawa_2022_Ross508}, the Calar Alto high-Resolution search for M dwarfs with Exoearths with Near-infrared
and optical Echelle Spectrographs \citep[CARMENES;][]{Ribas_2023_CARMENES_GTO}, and Habitable-zone Planet Finder \citep{Mahadevan_2012_HPF} and using transit photometers such as MEarth \citep{Charbonneau_2008_SPIE} and SPECULOOS \citep{Delrez_2018_SPECULOOS}. 
Nearby M dwarfs are also one of the main targets of former and ongoing space-based missions such as K2 \citep{2014PASP..126..398H} and TESS \citep{Ricker2015}. 

Transiting terrestrial planets around nearby M dwarfs are of particular importance in characterizing their atmospheres, which is already feasible through transmission and/or emission spectroscopy with current facilities.
So far, dozens of terrestrial-sized transiting planets have been discovered around M dwarfs within 30\,pc \citep[e.g.,][]{2017Natur.542..456G,2019Natur.573...87K, 2022ApJ...937L..17C}.
However, they typically orbit very close to their host stars. 
Their atmospheres are therefore vulnerable to the strong X-ray and ultraviolet (XUV) irradiation of their M dwarf hosts, which are known to be magnetically active in general. 
Indeed, recent spectroscopic emission observations of hot Earth-sized planets have refuted the existence of thick atmospheres around LHS\,3844b \citep{2019Natur.573...87K} and GJ\,1252b \citep{2022ApJ...937L..17C}, suggesting that
the atmospheres (if any) of such short-period planets have difficulty surviving. 

Currently, TRAPPIST-1 \citep{2017Natur.542..456G} is the only known M dwarf within 30\,pc transited by terrestrial planets with insolation of less than twice that of the Earth.
Due to the proximity (12\,pc) and small size (0.12\,$R_\odot$) of the host star \citep{2017Natur.542..456G}, the seven planets orbiting TRAPPIST-1 are the best known targets for characterizing atmospheres of warm-to-cool rocky planets with Earth or sub-Earth masses. 
However, it is unlikely that all of the planets harbor detectable atmospheres because their orbits are all close \citep[0.012--0.062\,au;][]{Agol_2021_TRAPPIST1} to the moderately active host star, whose X-ray-to-bolometric flux ratio is $\log L_{\rm X}/L_{\rm bol} \approx -3.5$ \citep{2017MNRAS.465L..74W}.
Recent secondary eclipse observations of TRAPPIST-1\,b and TRAPPIST-1\,c using the MIRI instrument on board the James Webb Space Telescope (JWST) suggest that they have little to no atmosphere, though a thin atmosphere of TRAPPIST-1\,c is not conclusively ruled out \citep{2023Natur.618...39G,Lincowski_2023}.  
The flat transmission spectra of the TRAPPIST-1 planets across the near-infrared wavelengths observed from the ground \citep{2019MNRAS.487.1634B} and space \citep{2016Natur.537...69D,2018NatAs...2..214D,2022A&A...658A.133G} are consistent with the absence of massive atmospheres.
The current list of low-insolation terrestrial planets amenable for atmospheric study is short, therefore it is essential to expand the sample of such planets to understand the diversity of atmospheres of temperate terrestrial planets.

In this paper, we present the discovery of a nearly Earth-sized planet with an orbital period of 12.76 days transiting the nearby (12\,pc), inactive M dwarf \target. 
The planet is one of the best temperate\footnote{We define temperate as an insolation between 4 and 0.25 times the Earth's insolation, following \citet{Triaud_2023_NatureAstro} and \citet{deWit_2024_JWSTRoadmap}, with the caveat that it is challenging to estimate the true surface temperatures of any planets without knowing their albedo and atmospheric composition and structure.} Earth-sized planets amenable for transmission spectroscopy, joining the planets around TRAPPIST-1.
The transit signal of this planet was first identified in the light curve data of TESS, and subsequently followed up by multiple ground-based observations.
\\

\section{TESS observations} \label{sec:tess_obs}

\begin{figure}[]
    \centering
    \includegraphics[width=0.4\textwidth]{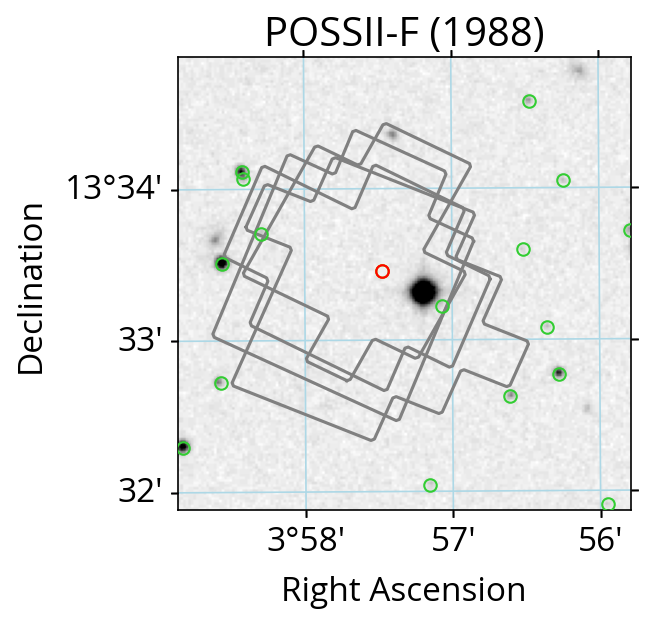}
    \caption{TESS photometric apertures from Sectors 42, 43, 57, and 70 (gray outlines) with an archival image from the POSSII-F survey taken in 1988 \citep{1991PASP..103..661R}, with Gaia DR2 source positions \citep{Gaia_2018_DR2_main} shown as circles; the red circle denotes the Gaia position (at epoch J2015.5) of \target, which is offset from the position in the archival image due to the star's proper motion, and the green circles show other sources.}
    \label{fig:tess-aper}
\end{figure}

\begin{figure*}[]
\centering
\includegraphics[width=16cm]{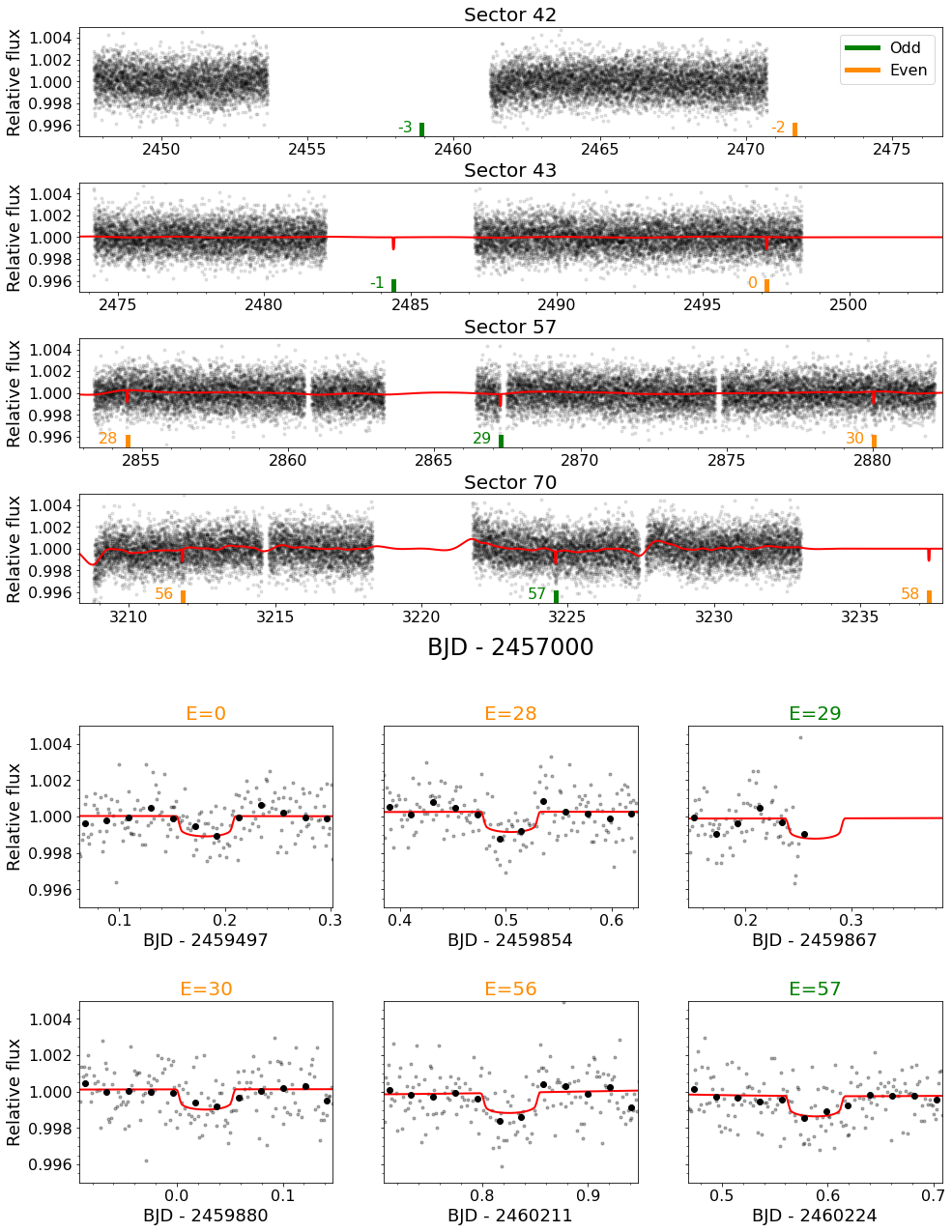}
\caption{(First–fourth rows) TESS PDCSAP light curves of \target from Sectors 42, 43, 57, and 70, respectively. The locations of the predicted transits assuming $P=12.76$\,days are marked by vertical bars, where green and orange indicate odd- and even-numbered transits, respectively. The relative transit epochs are labeled beside the bars. The red solid lines shown in the second–fourth panels are the light curves predicted by the maximum-likelihood transit+GP model. (Bottom two rows) Same as above, but the data around each of the six transits covered by TESS are zoomed in. The transit epochs are shown on top of each panel. The gray dots and black circles are the unbinned and 30-minute-binned data, respectively. The error bars of the binned data are shown, but they are smaller than the markers.
\label{fig:lc_TESS}}
\end{figure*}

TESS observed \target (TOI-6251) at a 2 minutes cadence in Sectors 42, 43, 57, and 70. The observations were conducted from UT 2021 August 20 to UT 2023 October 16, resulting in photometry spanning 797 days, with gaps for data downlink in the middle of each sector and gaps of 353 and 326 days between Sectors 43--57 and 57--70, respectively. Light curves were produced by the photometry pipeline developed by Science Processing Operations Center (SPOC) at NASA Ames Research Center \citep{Jenkins2016} using the apertures shown in 
Figure~\ref{fig:tess-aper}.
A transit signal with a period of 12.76 days was initially detected by the SPOC using an adaptive, noise-compensating matched filter \citep{2002ApJ...575..493J,2020TPSkdph} based on the data from Sectors 42, 43, and 57. An initial transit model including the effect of limb darkening was fitted to the signal \citep{Li:DVmodelFit2019}, and it passed a suite of diagnostic tests performed to assess its credibility \citep{2018PASP..130f4502T}, including the difference image centroiding test, which located \target within 9\farcs9 $\pm$ 8\farcs5 of the transit source.  
The TESS Science Office at Massachusetts Institute of Technology reviewed the Data Validation (DV) reports and issued an alert for this candidate planetary signal as TOI-6251.01 on UT 2023 April 3 \citep{2021ApJS..254...39G}. 
However, the orbital period was reported to be 25.52 days with a caution that the true orbital period may be half.
This was because only one odd-numbered\footnote{The transit number is counted such that the first transit event observed in the TESS data, namely Sector 43, is zero.} transit with respect to the 12.76 day period had been observed with TESS by the time the candidate was announced, and it occurred on the edge of a data gap in Sector 57 (see Figure~\ref{fig:lc_TESS}). As a result, the signal from odd-numbered transits was tentative.
This motivated us to schedule follow-up photometry for odd-numbered transits using MuSCAT2 and MuSCAT3 (Section~\ref{sec:muscat}), allowing us to confirm the true orbital period to be 12.76 days (Section~\ref{sec:LC_MuSCATs}). 
Subsequently, TESS Sector 70 data revealed an additional odd-numbered transit that also confirmed this period. 
We hereafter refer to this planet candidate as \planet (its planetary nature is validated in Section \ref{sec: FPScenario}).

For the subsequent analyses in this paper, we downloaded the Presearch Data Conditioning Simple Aperture Photometry \citep[PDCSAP;][]{Smoth_2012,Stumpe2012,2014PASP..126..100S} light curves of all available TESS sectors from the Mikulski Archive for Space Telescopes at the Space Telescope Science Institute, which are shown in Figure~\ref{fig:lc_TESS}.\\

\section{Follow-up Observations} \label{sec:ground_obs}

This section provides the details of ground-based observations to validate the planetary nature of the transit signal identified by TESS. 
The ground-based observations include adaptive optics (AO) imaging performed with Altair/NIRI and PyWFS/NIRC2 (Section \ref{sec:imaging}), multiband photometry with MuSCAT2 and MuSCAT3 \citep[][Section \ref{sec:muscat}]{Narita_2019_MuSCAT,Narita_2020_MuSCAT}, and radial velocity (RV) measurements with IRD and CARMENES (Section \ref{sec:doppler}).
We used data obtained by the IRD Strategic Survey Program of the Subaru Telescope (IRD-SSP), which independently selected \target as one of more than 100 M-type dwarfs to search for exoplanets using the RV technique \citep{Harakawa_2022_Ross508}.
Although the RV signal of \planet is too small to be detected by the IRD data alone, the RV measurements from IRD are used to rule out false-positive (FP) scenarios for the transit signal detected by TESS and constrain the presence of additional planets (Section \ref{sec:IRD_completeness}). These data, combined with the RV data from CARMENES, enable us to place an upper limit on the mass of \planet (Section \ref{sec:joint}).

\subsection{AO Imaging}  \label{sec:imaging}

\target was observed with the NIRI imager of the Gemini-North telescope \citep{Hodapp_2003_NIRI} with the AO instrument Altair \citep{Herriot_2000_Altair} in the program of GN-2009B-Q-10-213 (PI: S. Dieterich).
The observations were conducted on UT 2009 September 18 using the near-infrared band filters at $J$ ($\approx$1.2\,$\mu$m), $H$ ($\approx$1.6\,$\mu$m), and $K$ ($\approx$2.2\,$\mu$m).   
We reduced the Gemini/NIRI data downloaded from the Gemini data archive. 
The images after the data processing are shown in Figure \ref{fig:AOimage} with the 5$\sigma$ contrast curves at $J$, $H$, and $K$.
See Appendix \ref{app: highcontrast_detail} for details about the observations and data reduction.  
As shown in Figure \ref{fig:AOimage}, we found that there is a possible faint source only in the $J$-band image, which is located at a projected separation of $\approx$2\farcs{4} from \target. \par
We derived contrast limits from the $J$, $H$, and $K$ images to assess how likely it is that the identified transit signal is an FP (see Section \ref{sec: FPScenario}). To that end, we first subtracted the radial profiles of the central star's point spread function (PSF). 
We then convolved the images with circular apertures whose diameters are the same as the full-width-at-half-maximum (FWHM) of the central star's PSFs and computed standard deviations as a function of projected separations from \target in the convolved images. 
We calculated the central star's flux in the apertures used at the image convolutions and divided the values of the standard deviations by the fluxes of the central star. 
As a result, we obtained the 5$\sigma$ contrast limits shown in Figure \ref{fig:AOimage}.
The $J$-, $H$-, and $K$-band contrast limits at 1\farcs{0} are 8.1, 8.5, and 9.0\,mag, respectively. 
The source detected at $\approx$2\farcs{4} from \target has a contrast equal to 9.23--9.27\,mag depending on linearity corrections in the $J$-band master image (see Appendix \ref{app:reduc_NIRI}), yielding the source's $J$-band magnitude as approximately 17.9\,mag.
The $K$-band contrast limit suggests that the faint source has a contrast greater than $\approx$9.8\,mag, indicating that the source is fainter than $\approx$17.6\,mag at $K$.
We then referred to \target's $J$- and $K$-band magnitudes in Table \ref{tab:stellar_para} to convert the contrast measurements of the source to the apparent magnitudes.  
We discuss in Section \ref{sec: FPScenario} that this faint object cannot be the source of the transit signal detected in the TESS data.

\begin{table}[]
\centering
\caption{Stellar parameters of \target}{
    \begin{tabular}{lcc}
        \hline
        \hline
        Parameter & Value & Reference\tablenotemark{a} \\
        \hline
        R.A. (J2000.0) & 00 15 49.242 & Gaia DR3 \\
        Decl. (J2000.0) & +13 33 22.32 & Gaia DR3 \\
        $\mu_{\mathrm{R.A.}}\cos{\delta}$ (mas yr$^{-1}$) & +618.065 $\pm$ 0.039 & Gaia DR3 \\
        $\mu_{\mathrm{decl.}}$ (mas yr$^{-1}$) & +329.446 $\pm$ 0.034 & Gaia DR3 \\
        $\pi$ (mas) & 82.1938 $\pm$ 0.0326 & Gaia DR3 \\
        Distance (pc) &  $12.1664 \pm 0.0049$ & Gaia DR3 \\
        RUWE & 1.198 & Gaia DR3 \\
        Gaia $G$ (mag) & 11.399 $\pm$ 0.003 & Gaia DR3 \\
        Gaia $B_{\rm{P}}$ (mag) & 12.831 $\pm$ 0.003 & Gaia DR3 \\
        $T$ (mag) & 10.177 $\pm$ 0.007 & TIC \\
        $J$ (mag) & 8.619 $\pm$ 0.020 & 2MASS \\
        $K_{\rm{s}}$ (mag) & 7.807 $\pm$ 0.020 & 2MASS \\
        Spectral type & M3.0\,V & Rei95 \\
        $T_{\rm{eff}}$ (K) & $3296\ ^{+48}_{-36}$  & This work \\
        log $g$ (cgs) & $5.21 \pm 0.07$ & This work \\
        $M_{\rm s}$ ($M_{\rm{\odot}}$) &  $0.2414 \pm 0.0060$ & This work \\
        $R_{\rm s}$ ($R_{\rm{\odot}}$) &  $0.2617\ ^{+0.0058}_{-0.0070}$ & This work \\
        $\rho_{\rm s}$ ($\rho_\odot$)  &  $13.5\ ^{+1.2}_{-0.9}$ & This work \\
        $L_{\rm bol}$ (10$^{-3}$ $L_{\rm{\odot}}$) &  $7.28 \pm 0.15$ & This work \\
        $L_{\rm X}$ (10$^{25}$\,erg\,s$^{-1}$)&  $6.0 \pm 1.3$ & This work\\
        $\log{R'_{\rm HK}} $ & $-5.23^{+0.10}_{-0.14}$ & This work\\
        $P_{\rm{rot}}$ (days) &  85 & This work  \\
        $v \sin{i}$ (km s$^{-1}$) & $<$ 2 & This work \\   
        $[\rm{Fe/H}]$ & $-0.32 \pm 0.06$ & This work\\
        $[\rm{Na/H}]$ & $-$0.37 $\pm$ 0.11 & This work\\
        $[$Mg/H$]$ & $-$0.32 $\pm$ 0.22 & This work\\
        $[$Ca/H$]$ & $-$0.37 $\pm$ 0.12 & This work\\
        $[$Ti/H$]$ & $-$0.26 $\pm$ 0.26 & This work\\
        $[$Cr/H$]$ & $-$0.34 $\pm$ 0.15 & This work\\
        $[$Mn/H$]$ & $-$0.41 $\pm$ 0.19 & This work\\
        $[$Sr/H$]$ & $-$0.22 $\pm$ 0.20 & This work\\
        \hline
    \end{tabular}}
    \tablenotetext{a}{Gaia DR3: \citet{2023A&A...674A...1G}; TIC: \citet{Paegert_2021_TIC}; 2MASS: \citet{Skrutskie_2006_2MASS}; 
     Rei95: \citet{Reid_1995}.}
    \label{tab:stellar_para}
\end{table}

On UT 2023 August 15, we also obtained high spatial-resolution images of the \target system using NIRC2 and its $K^{\prime}$ ($\approx$2.12\,$\mu$m) filter on the Keck II Telescope (PI: C. Beichman).
The images of \target were corrected by the AO system for NIRC2, where we adopted PyWFS \citep{Bond2020_PyWFS_Keck} as a wavefront sensor.
PyWFS performs wavefront sensing at near-infrared wavelengths (mainly $H$ band).
Thus, PyWFS is an ideal choice to obtain high-quality PSF corrections for objects that are faint in the visual but bright in the infrared such as M-type stars \citep{Uyama_2023_IRDCompanion}.
The Keck/NIRC2 $K^{\prime}$-band image after data processing is shown in Figure \ref{fig:AOimage} with the 5$\sigma$ contrast curve at $K^{\prime}$.
In the image, we detect no object and did not recover the detection of the faint source identified in the Gemini/NIRI $J$-band image (see above).
Appendix \ref{app:reduc_NIRC2} presents more details on our Keck/NIRC2 observations and imaging data reduction.  \par

As done for the Gemini/NIRI images, we derived a contrast limit from the Keck/NIRC2 image for FP validation (see Section \ref{sec: FPScenario}).   
We followed the same procedures adopted for the Gemini/NIRI observations except that the diameter of the circular aperture used for image convolution was set equal to $\lambda/D$ (where $D$ is the diameter of the Keck telescope)\footnote{This is because the FWHM of the unsaturated Br$_{\mathrm{\gamma}}$ PSF was measured to be smaller than $\lambda/D$.}.
Figure \ref{fig:AOimage} shows the 5$\sigma$ contrast limit from the Keck/NIRC2 observations. 
The $K^{\prime}$-band contrast limit is 9.0\,mag at 1$^{\prime\prime}$ and slightly deeper ($\sim$0.2\,mag on average) than the $K$-band contrast limit from Gemini/NIRI over 0\farcs{1} to 2\farcs{8}.

\begin{figure*}[]
\centering
\includegraphics[scale=0.086,trim=5mm 0mm 0mm 0mm,clip]{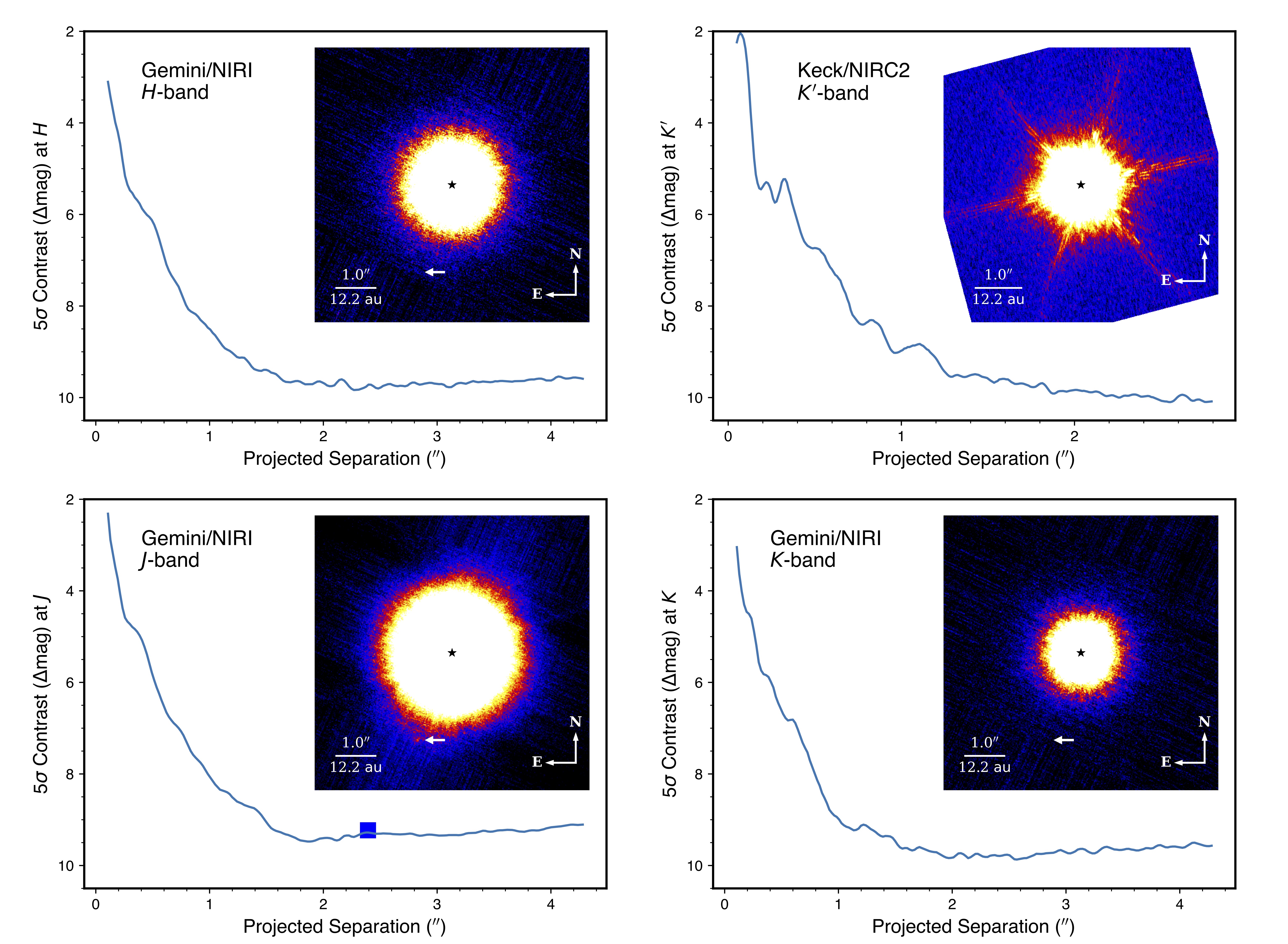}
\caption{5$\sigma$ contrast limits at $J$ (bottom left), $H$ (top left), and $K$ (bottom right) from Gemini/NIRI and at $K^{\prime}$ from Keck II/NIRC2 (top right). 
The combined images are shown in the insets of the panels, where \target is at the star symbols.  The maximum values of the image dynamic range were set to be 20 times higher than the noise levels at the edge areas.  
The locations corresponding to the identified possible source that was detected only in the $J$-band image are marked by arrow symbols and its contrast by a square symbol. 
Note that the source detection and contrast measurement were performed with the images after PSF subtractions.
\label{fig:AOimage}}
\end{figure*}

\subsection{Multiband Transit Photometry} \label{sec:muscat}

We conducted photometric observations for four transits of \planet that were expected to occur only if the true orbital period is 12.76 days (i.e., odd-numbered transits) using the multiband imagers MuSCAT2 \citep{Narita_2019_MuSCAT} on the 1.52\,m Telescopio Carlos S\'anchez at the Teide Observatory in the Canary Islands, Spain, and MuSCAT3 \citep{Narita_2020_MuSCAT} on the 2\,m Faulkes Telescope North (FTN) at the Haleakal\={a} Observatory in Hawai'i, USA, on UT 2023 July 22 (MuSCAT3), 2023 August 16 (MuSCAT2), 2023 September 11 (MuSCAT3), and 2023 October 6 (MuSCAT2).
Both MuSCAT2 and MuSCAT3 have four optical channels that enable simultaneous imaging in the $g$, $r$, $i$, and $z_{\rm s}$ bands.
Each channel of MuSCAT2 is equipped with a 1k $\times$ 1k pixel CCD with a pixel scale of 0\farcs{44}, providing a field of view (FoV) of 7\farcm{4}.
On the other hand, each channel of MuSCAT3 is equipped with a 2k $\times$ 2k pixel CCD with a pixel scale of 0\farcs{27}, providing an FoV of 9\farcm{1}.
During the observations, we defocused the telescopes to avoid saturation such that the FWHM of the stellar PSF was 4\farcs{5}--5\farcs{5},  7\farcs{0}--8\farcs{0}, 4\farcs{0}--5\farcs{0}, and 8\farcs{5}--10\farcs{0} for the July 22, August 16, September 11, and October 6 observations, respectively. 
The exposure times were set at 7--60\,s depending on the instrument, filter, and night. 

After calibrating the obtained images for dark and flat field, we performed aperture photometry using the pipeline described in \citet{2011PASJ...63..287F}, with optimal aperture radii of 18--24 pixels (7\farcs8--10\farcs4) for the MuSCAT2 data and 15--19 pixels (4\farcs1--5\farcs1) for the MuSCAT3 data, depending on the band and night. We note that the photometric dispersions of the $g$-band data in all nights are much larger than the expected transit depth ($\sim$0.1\%), and we therefore discard these data for the subsequent analyses of this paper.
The derived light curves in the $r$, $i$, and $z_s$ bands from the four nights {\footnote{These time-series photometric data can be provided upon request to the first authors.} are shown in Figure \ref{fig:lc_MuSCATs}.

\begin{figure*}[]
\centering
\includegraphics[width=16cm]{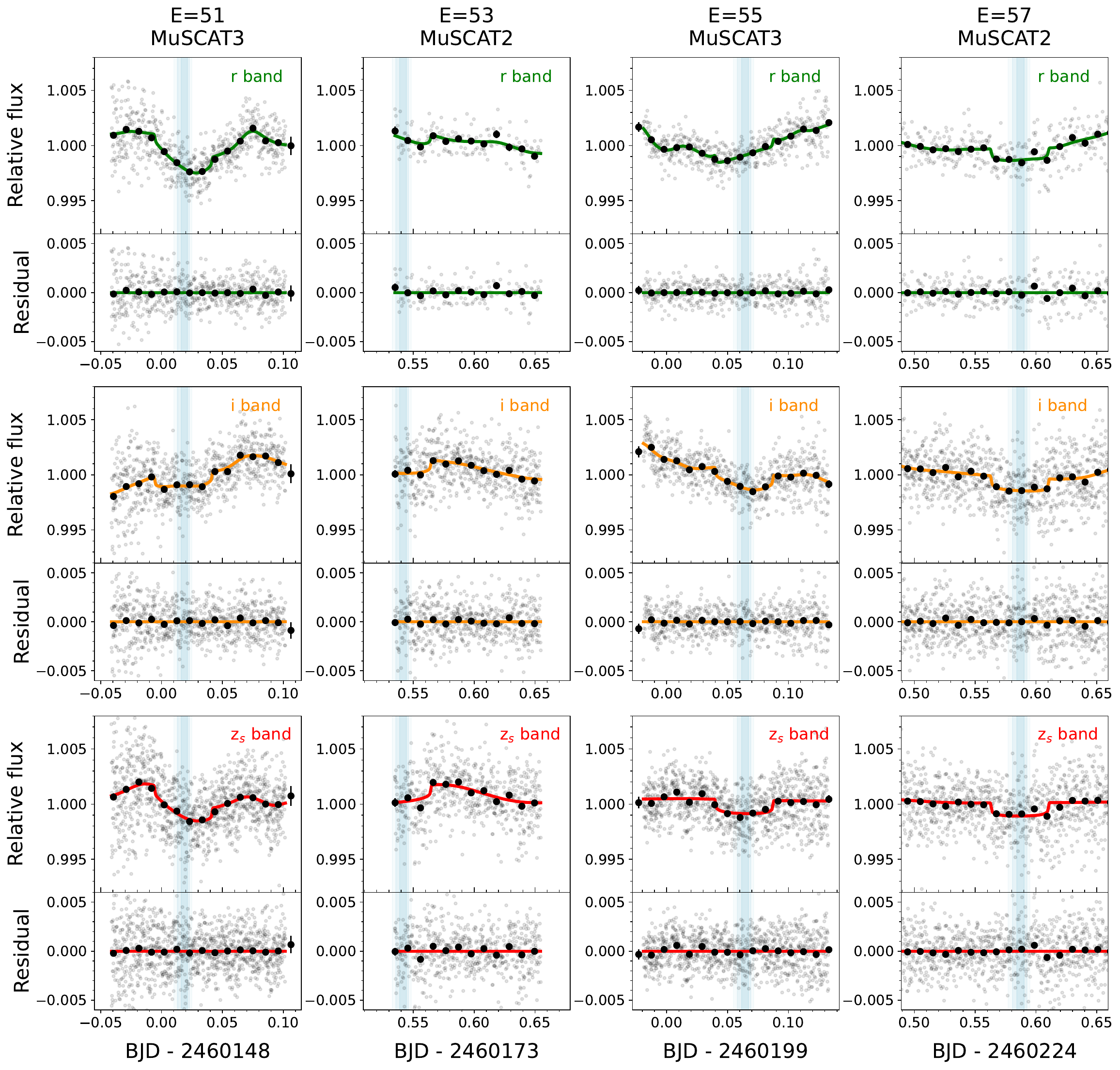}
\caption{
Individual transit light curves of \planet obtained with MuSCAT2 and MuSCAT3. Columns from left to right show the data taken on the nights of UT 2023 July 22 (transit epoch = 51), August 16 (53), September 11 (55), and October 6 (57). The top, middle, and bottom {rows} are for the $r$, $i$, and $z_s$ bands, respectively. The top and bottom panels in each panel pair show the undetrended light curve and its residuals from the best-fit transit+GP model (colored line), respectively (see details in Section \ref{sec:LC_MuSCATs}). For both panels, gray dots and black circles indicate unbinned and 15 minute binned data, respectively. Darker to lighter blue vertical bands indicate the 1$\sigma$, 2$\sigma$, and 3$\sigma$ credible regions, respectively, of the midtransit times predicted from $P$ and $T_{c,0}$ derived from the analysis of the TESS light curves (Section \ref{sec:LC_TESS}).
\label{fig:lc_MuSCATs}}
\end{figure*}

\subsection{Doppler Spectroscopy} \label{sec:doppler}

We observed \target using the IRD instrument \citep{2018SPIE10702E..11K} on the Subaru Telescope at Maunakea Observatory in Hawai'i, USA, between UT 2019 June 22 and UT 2022 November 25. 
\target is in the master target list of the IRD-SSP campaign and has been intensively observed after initial screening observations \citep{Harakawa_2022_Ross508}.
In IRD observations, wavefront-corrected object light is transferred to IRD's temperature-stabilized spectrograph in a vacuum chamber by a multimode fiber and dispersed by the spectrograph over the wavelength range of $\approx$930--1740\,nm with the resolving power ($\mathcal R$) of $\approx$70,000 \citep{2018SPIE10702E..11K}.  
Using another multi-mode fiber, we obtained laser frequency comb spectra simultaneously with the target spectra, enabling us to calibrate RV drifts and instrument profile fluctuations \citep{Hirano_2020_IRDPipeline}. \par
We took 76 spectra by spending exposure times of 280--1100\,s for each data acquisition, with a signal-to-noise ratio (S/N) ranging from 20 to 158 around 1000\,nm.
We extracted one-dimensional spectra from the raw two-dimensional images taken with two H2RG (HAWAII-2RG) detectors of the IRD spectrograph based on the same procedures as used in the previous IRD papers \citep[e.g.,][]{
Kuzuhara_2018_SPIE,Hirano_2020_IRDPipeline,Harakawa_2022_Ross508}. \par
We applied the pipeline of \citet{Hirano_2020_IRDPipeline} to the 1D spectra, providing the RV measurements of \target reported in Appendix \ref{sec: RV_measures}.
The pipeline divides a spectrum into many segments and calculates RVs from each segment, which are statistically combined into a single RV for each exposure. 
In this work, we adopted more constraining criteria than those adopted by \citet{Hirano_2020_IRDPipeline} when clipping outlier RV measurements in the statistical combination of RVs, providing more stable RV measurements but slightly enlarging the internal errors.  
The zero-points of the IRD RV measurements were corrected  using the postprocessing described by \citet{Gorrini_2023_CARMENES+IRD}. 
The internal error of RV measurements is typically 3.7\,m\,s$^{-1}$.

We also observed \target with CARMENES \citep{Quirrenbach_2014_CARMENES} in the framework of the CARMENES-TESS project for TESS follow-up of temperate planets (PI: E.\,Palle).
CARMENES is a fiber-fed, dual-channel spectrograph at the 3.5 m Calar Alto telescope that operates simultaneously in the wavelength range between 0.52\,$\mu$m and 1.71\,$\mu$m in one shot.
A dichroic at 0.96\,$\mu$m splits the light that is injected into the two spectrograph channels, namely, VIS in the optical ($\mathcal R \approx$ 94,600) and NIR in the near-infrared ($\mathcal R \approx$ 80,400).
We used the same standard operational mode as in the guaranteed time observations program for planet RV search \citep{Ribas_2023_CARMENES_GTO}, in which target spectra were obtained simultaneously with the spectra from a Fabry-P\'erot etalon for wavelength calibration using variable exposure times limited by S/N = 150 and efficient scheduling \citep{Garcia-Piquer_2017}.
We channeled the CARMENES data through our pipelines, which include spectrum extraction and nightly zero-point corrections \citep[][and references therein]{Ribas_2023_CARMENES_GTO}, telluric absorption corrections \citep{Nagel_2023_CARMENES}, 
and RV determination with {\tt serval} \citep{Zechmeister_2018_SERVAL}.
This work adopts the RVs measured with the CARMENES VIS channel, which are given in Appendix \ref{sec: RV_measures}.
Because of the larger number of telluric-free lines in the optical spectrum of a mid-M dwarf with respect to the near-infrared \citep{Reiners_2018_CARMENES}, the RVs from the CARMENES VIS channel have lower internal errors than IRD, of about 1.7\,m\,s$^{-1}$.

\subsection{X-ray Flux Measurement}\label{sec:xray}
The European Space Agency X-ray telescope XMM-Newton observed \target on UT 2018 June 16 (PI: J.\,H.\,M.\,M. Schmitt). 
XMM-Newton \citep{Jansen2001}  
simultaneously operates two high-spectral resolution detectors (Reflection Grating
Spectrometer) and an optical monitor, both of which did not attain a signal of \target, and three European Photon Imaging Camera (EPIC-pn and EPIC-MOS) detectors \citep[sensitivity range 0.1--15\,keV and 0.2--10\,keV, respectively, and spectral resolution $E / \Delta E \sim$ 20--50;][]{Strueder2001,Turner01}.
A $4.5\sigma$ detection of the star was achieved with an average 17.8\,ks exposure on the three EPIC detectors. 
The X-ray spectra were analyzed using standard tools, revealing a typical one-temperature coronal spectrum at $\log T ({\rm K}) = 6.44^{+0.12}_{-0.11}$ with an emission measure of $\log {\rm EM~(cm}^{-3}$)$=48.39^{+0.14}_{-0.19}$. 
The fit assumed photospheric metallicity (see Section~\ref{sec:stellar_param}) and a small interstellar medium absorption of $N_{\rm H}=3 \times 10^{18}$\,cm$^{-3}$. 
The spectral fit resulted in an X-ray (range 5--100\,\AA) luminosity of $L_{\rm X} = 6.0 \pm 1.3 \times 10^{25}$\,erg\,s$^{-1}$. 

\section{Analysis and Results}\label{Ana_Res}

\subsection{Stellar Parameters}\label{sec:stellar_param}

Using a template spectrum created from the IRD spectra following \citet{Hirano_2020_IRDPipeline}, we carried out a line-by-line analysis to determine the effective temperature ($T_{\rm{eff}}$), metallicity ([Fe/H]), and chemical abundances of \target.
We followed the method developed by \citet{Ishikawa_2020}, which measures the equivalent widths (EWs) of the molecular lines and some atomic lines and compares them with those from theoretical synthetic spectra.  
We focused on the molecular lines of FeH and the atomic lines of Na, Mg, Ca, Ti, Cr, Mn, Fe, and Sr via the procedures adopted in the previous IRD-SSP papers \citep[see][for details]{Harakawa_2022_Ross508,Ishikawa_2022_abundance,Hirano_2023_K2}. 
The $T_{\rm{eff}}$ and [Fe/H] of \target were derived to be $3344 \pm 44$\,K and $-0.43 \pm 0.15$\,dex, respectively. 
Table \ref{tab:stellar_para} provides the abundances of the other atomic elements, which are available to characterize the \target system in the future. 
We also analyzed the high-S/N CARMENES VIS and NIR template spectra of \target using the spectral synthesis code {\tt SteParSyn} \citep{Tabernero_2022}, deriving \target's $T_{\rm eff}$, $\log{g}$, 
and [Fe/H] to be 3363 $\pm$ 70\,K, 5.21 $\pm$ 0.07\,dex, 
and $-$0.29 $\pm$ 0.07\,dex, respectively. 
The $T_{\rm eff}$ and [Fe/H] values from the CARMENES spectra are consistent with those from the IRD spectra within their error bars.
We computed the weighted average of the two [Fe/H] measurements to be $-0.32 \pm 0.06$ dex, which is given in Table \ref{tab:stellar_para}.
The relatively low metallicity agrees with the previous estimates from spectroscopic observations of \target, which range from $-$0.29 $\pm$ 0.09 to $-$0.17 $\pm$ 0.13\,dex 
\citep{Newton_2014, 2020A&A...644A..68M}. 
\citet{2016ApJ...822...97H} also inferred a lower metallicity of $-$0.51$\pm$ 0.15\,dex from their metallicity--radius empirical correlation for M3V stars.
We used the CARMENES spectra to constrain the projected rotational velocity of \target to be $v \sin{i} <$ 2\,km\,s$^{-1}$ \citep{Reiners_2018_CARMENES}. \par

The metallicity of \target is lower than the average value ([Fe/H] $= -0.05$\,dex) of the Galactic thin disk \citep{2017MNRAS.464.2610F}.
However, the following kinematic analysis supports that this object belongs to the Galactic thin disk.
We calculated the space velocities $UVW$ of \target relative to the local standard of rest (LSR) based on the astrometric measurements in Gaia DR3 \citep{2023A&A...674A...1G} and the absolute RV \citep{2018A&A...616A...7S}.
Then, we adopted the distance of 8.2\,kpc between the Sun and the Galactic center \citep{2017MNRAS.465...76M}, the height of the Sun above the Galactic midplane of 0.025\,kpc \citep{2008ApJ...673..864J}, and the solar motion relative to the LSR of ($U_\odot$, $V_\odot$, $W_\odot$) = (11.10, 12.24, 7.25)\,km\,s$^{-1}$ \citep{2010MNRAS.403.1829S}.
The calculated $U$, $V$, and $W$ are $-$39.88\,km\,s$^{-1}$, 38.92\,km\,s$^{-1}$, and $-$23.47\,km\,s$^{-1}$, respectively.
These kinematics enabled us to compute the relative probability for the thick-disk-to-thin-disk membership ($TD/D$) using the method of \citet{2014A&A...562A..71B}.
The computed $TD/D$ is 0.16, which meets the criteria of $TD/D < 0.5$ to be classified as a likely thin disk star.

We also estimated the mass of \target to be $M_s = 0.2414 \pm 0.0060$\,$M_{\rm{\odot}}$ using the $K_s$ band magnitude from the Two Micron All Sky Survey \citep[2MASS; ][]{Skrutskie_2006_2MASS}, the distance $d$ (= $12.1664 \pm 0.0049$\,pc) derived from a parallax measurement for \target in Gaia DR3, and an empirical mass-$M_K$ relation for M dwarfs of 
\citep[][their Equation (4) with $n = 5$]{Mann_2019_Mass}.
To assign an error to the mass estimation, we performed a Monte Carlo simulation adopted by \citet{Harakawa_2022_Ross508}.

The $T_{\rm{eff}}$, bolometric luminosity ($L_{\rm bol}$), and radius ($R_{\rm{s}}$) of \target were determined by analyzing the spectral energy distribution (SED) created from the photometric measurements in the $G$, $B_P$, and $R_P$ bands from Gaia DR3; the $J$, $H$, and $K_s$ bands from 2MASS; and the $W1$, $W2$, $W3$, and $W4$ bands from AllWISE \citep{2010AJ....140.1868W,2011ApJ...731...53M}.
The SED is modeled by synthetic spectra using the BT-Settl grid stellar atmospheric models of \citet{Allard_2015_BTSettl}, following the procedure described by \citet{Harakawa_2022_Ross508}.  
We employed a Markov Chain Monte Carlo (MCMC) simulation using \texttt{emcee} \citep{2013PASP..125..306F} for the model parameters of $T_{\rm eff}$, $R_{\rm s}$, $M_{\rm s}$, [Fe/H], and $d$, imposing Gaussian priors on $M_{\rm s}$, [Fe/H], and $d$ using the values listed in Table\,\ref{tab:stellar_para}. Note that we expand the widths of the priors for $M_{\rm s}$ and [Fe/H] by a factor of 5 to take into account the possible systematic uncertainties in the stellar atmospheric model grid.
As a result, we obtained $R_{\rm s} = 0.2617^{+0.0057}_{-0.0070}\,R_\odot$ and $T_{\rm eff} = 3296^{+48}_{-36}$\,K.
From $R_{\rm s}$, $T_{\rm eff}$, and the Stefan-Boltzmann law, $L_{\rm bol}$ was calculated to be $7.28 \pm 0.15 \times 10^{-3}\,L_\odot$.
We summarize the derived stellar parameter values in Table~\ref{tab:stellar_para}. 
Note that the derived $T_{\rm{eff}}$ from the SED analysis is consistent with the spectroscopically derived ones ($3344 \pm 44$\,K from IRD and $3363 \pm 70$\,K from CARMENES) within $\approx$1$\sigma$. We adopted the $T_{\rm{eff}}$ from the SED analysis for the rest of the paper.

\subsection{Confirmation of the Transit Signal}

As described in Section \ref{sec:tess_obs}, a transit signal with an orbital period of 12.76 days was initially detected in the TESS data from Sectors 42, 43, and 57, and was later reinforced by additional data from Sector 70. The multiple event detection statistic (MES) of the transit signal for all the available sectors, calculated by the SPOC pipeline, is 9.5, which is well above the detection threshold of MES~=~7.1. However, although the transit depths individually measured for odd- and even-numbered transits are consistent with each other, the significance of the transit signal from the odd-numbered transits alone is only $\sim$1$\sigma$ (see Figure \ref{fig:odd_even} in Appendix \ref{sec:odd_even}). 
Therefore, follow-up observations can confirm the 12.76 day transit signal by increasing the detection significance of odd-number transits. Confirming the transit signal by follow-up observations with spatial resolutions higher than TESS is also helpful to rule out the possibility that the transit signal detected by TESS comes from any contaminating stars within the TESS aperture.
To this end, we separately analyze the TESS (Section \ref{sec:LC_TESS}) and ground-based (Section \ref{sec:LC_MuSCATs}) data, the latter of which come only from odd-numbered transits, and compare the consistency in the transit model parameters between the two data sets.

\subsubsection{Analysis of the TESS Light Curves}
\label{sec:LC_TESS}

We first derived the transit model parameters from the TESS data alone by simultaneously modeling the TESS PDCSAP light curves of Sectors 43, 57, and 70 with a transit and systematics model. For the transit, we adopted the Mandel \& Agol model implemented by {\tt PyTransit} \citep{Parviainen2015} with the following parameters: log-scaled semimajor axis $\ln(a_R)$, where $a_R \equiv a/R_s$ and $a$ is semimajor axis; impact parameter $b$; planet-to-star radius ratio $k$, where $k \equiv R_p/R_s$ and $R_p$ is the planetary radius; orbital period $P$; reference midtransit time $T_{{\rm c},0}$; and two coefficients $u_1$ and $u_2$ of a quadratic limb-darkening law. We assumed a circular orbit; i.e., the eccentricity $e$ was fixed at 0.

Simultaneously with the transit model, we also estimated the time-correlated systematic noise in the light curves using a Gaussian process (GP) model implemented in {\tt celerite} \citep{2017AJ....154..220F} with a kernel function of a stochastically-driven, damped simple harmonic oscillator. For the GP, the modeled parameters were the frequency of undamped oscillation $\omega_0$, the scale factor to the amplitude of the kernel function $S_0$, and the quality factor $\mathcal{Q}$. We set $\mathcal{Q}$ to unity in all sectors and allowed $\omega_0$ and $S_0$ (as $\ln \omega_0$ and $\ln S_0$, respectively) to be free for each sector.

We ran an MCMC analysis using {\tt emcee} to estimate the posterior probability distributions of the parameters. 
For $u_1$ and $u_2$, we applied Gaussian priors with the values and uncertainties given by {\tt LDTk} \citep{Parviainen2015b} for the stellar parameters of \target in Table~\ref{tab:stellar_para}. Note that we enlarged the uncertainties of $u_1$ and $u_2$ provided by {\tt LDTk} by a factor of 3 taking into consideration the possible systematics in the stellar models. 
We applied uniform priors for the other parameters. We started the analysis with initial values of our best guess, including $P=12.76$\,days.
The priors and posteriors of the transit model parameters are summarized in Table~\ref{tab:lc_analysis} of Appendix \ref{sec:lc_analysis}.
Figure\,\ref{fig:corner_transit} shows the obtained posteriors of $k$, $b$, $\ln{(a_R)}$, $P$, and the transit duration $T_{14}$, which is calculated from the former four parameters. The model light curve for each sector predicted by the maximum-likelihood transit+GP model is shown in red in Figure \ref{fig:lc_TESS}.

\begin{figure}[]
\centering
\includegraphics[width=8cm]{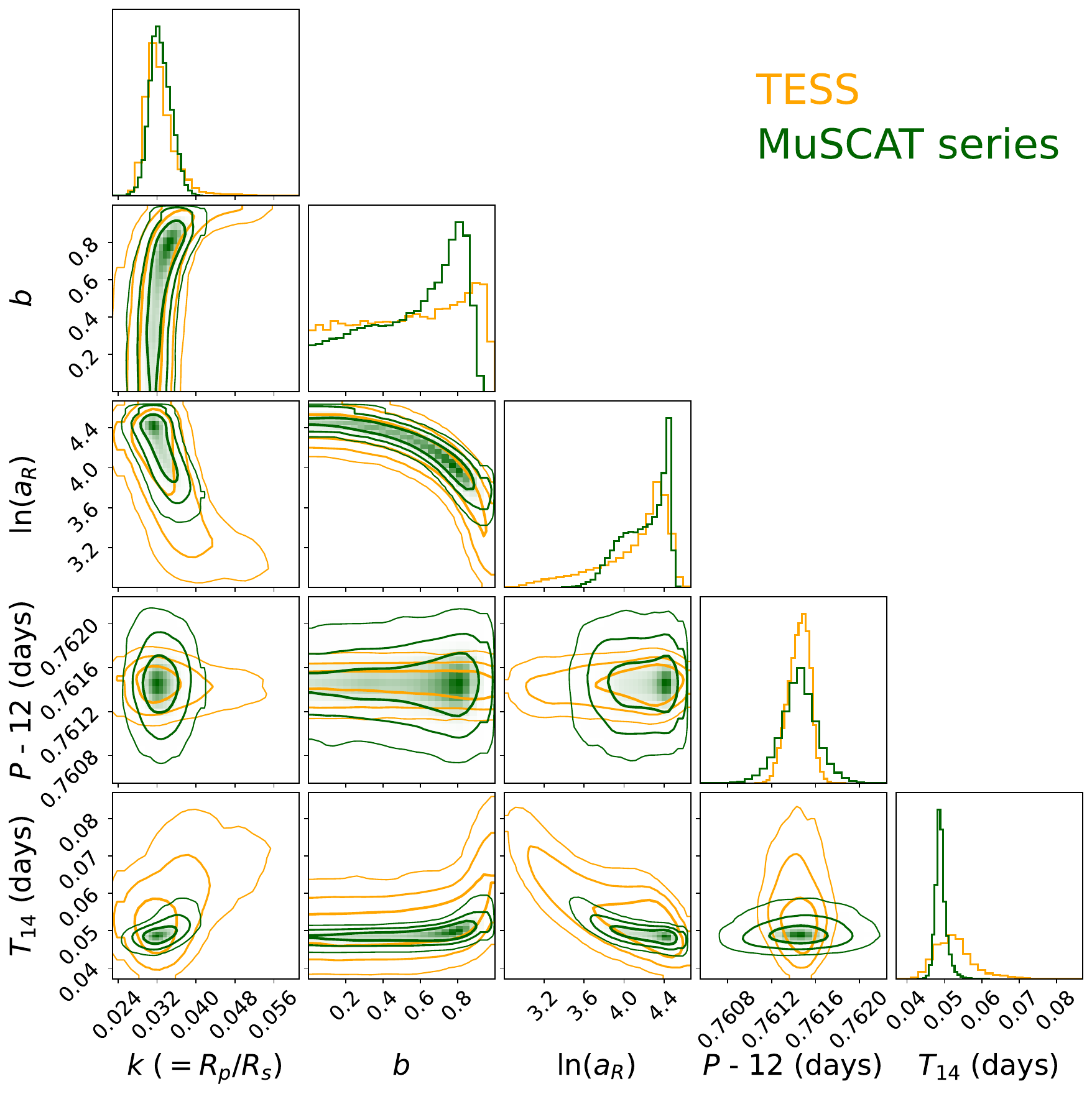}
\caption{
Corner plot for the posterior probability distributions of $k$, $b$, $\ln{(a_R)}$, $P$, and $T_{14}$ from the analyses on the TESS (orange; Section \ref{sec:LC_TESS}) and MuSCAT series (green; Section \ref{sec:LC_MuSCATs}) light curves, respectively. Thicker to thinner contours indicate the 1$\sigma$, 2$\sigma$, and 3$\sigma$ credible regions. The posterior distributions of all parameters are in agreement with each other between the two data sets.
\label{fig:corner_transit}}
\end{figure}

\subsubsection{Analysis of the Ground-based Light Curves} \label{sec:LC_MuSCATs} 

We analyzed the light curves obtained with MuSCAT2 and MuSCAT3 to confirm that ($i$) the transit signal detected by TESS indeed comes from \target with the higher spatial resolution of the MuSCAT series, and ($ii$) the true orbital period is 12.76\,days instead of twice it.
Because all the light curves obtained with MuSCAT2 and MuSCAT3 correspond to odd-numbered transits of \planet for the ephemeris of $P=12.76$\,days, while most of the transit signals in the TESS data come from even-numbered transits (four of the five full transits), the true orbital period can be confirmed by detecting or rejecting a transit signal in these MuSCAT2 and MuSCAT3 light curves. \par
Although a hint of transit signal is seen in each MuSCAT2 and MuSCAT3 light curve, the significance of each signal is marginal. We therefore simultaneously modeled all 12 MuSCAT2 and MuSCAT3 light curves with transit+GP models as follows. \par
For the transit, we applied the same model as for the analysis of the TESS light curves in Section \ref{sec:LC_TESS}. All model parameters are shared across all light curves except for $u_1$ and $u_2$, which were let free for each band.
For the GP model, we adopted an approximate Mat\'{e}rn 3/2 kernel implemented in the {\tt celerite} package with hyperparameters of $\ln \sigma$ and $\ln \rho$, where $\sigma$ and $\rho$ are the signal amplitude and length scale, respectively. We let $\ln \sigma$ be free for each light curve (each transit and each band) while letting $\ln \rho$ be shared within the three light curves of each transit assuming that the timescale of the time-correlated noise is common among all bands for a given night. 

We then performed an MCMC analysis using {\tt emcee} to calculate the posterior probability distributions of the model parameters. We applied uniform priors to all parameters except for $u_1$ and $u_2$, to which Gaussian priors were applied in the same way as for the analysis in Section \ref{sec:LC_TESS}. We limited the values of $\ln \rho$ to --3 or larger so that the timescale of systematic trends is larger than the duration of the transit ingress and egress. We started the analysis setting the values of $k$, $b$, $a_R$, $P$, and $T_{\rm c,0}$ to be the best-fit values obtained in the analysis of the TESS light curves. 
See Table~\ref{tab:lc_analysis} of Appendix \ref{sec:lc_analysis} for the priors and posteriors of the transit model parameters.

We detected a constant-period transit signal with a radius ratio of $k = 0.0326^{+0.0025}_{-0.0019}$ at 17$\sigma$ significance.
In Figure \ref{fig:corner_transit}, we show the derived posterior probability distributions of $k$, $b$, $\ln(a_R)$, $P$, and $T_{14}$ in green. The distributions are highly consistent with those from the analysis of the TESS data, which are shown in orange. This consistency indicates that there does exist a transit signal in the MuSCAT2 and MuSCAT3 data that is consistent with the transit signal detected by TESS. We therefore conclude that ($i$) the transit signal detected in TESS comes from the photometric aperture of MuSCAT2 and MuSCAT3 ($\lesssim 10''$ in radius) centering on \target, and ($ii$) the true orbital period is 12.76\,days.

\subsection{RV Periodogram and Stellar Activity} \label{sec:IRD_RV_Activity}

We performed the generalized Lomb Scargle (GLS) periodogram \citep{Zechmeister_2009} using a tool of \texttt{astropy} to search for a periodic signal in our RV measurements and activity indicators from IRD spectra, based on \citet{Harakawa_2022_Ross508}.
We did not analyze the RVs and indicators from CARMENES due to its limited observing span. 
The activity indicators that we used were FWHM, dV, CRX, and dLW.
The method to compute the indicators with IRD data is described further in \citet{Harakawa_2022_Ross508}.
In the following analyses, the false alarm probability (FAP) calculations are performed based on \citet{Baluev_2008}.
The FWHM and dV (or BiGauss) correspond to the FWHM and line asymmetry of the absorption lines, respectively. 
In addition, CRX and dLW correspond to the dependence of RV measurements on wavelength and differential line width, respectively.
See \citet{Zechmeister_2018_SERVAL} for the full definitions of the above activity indicators, except for dV, which is defined in \citet{Santerne_2015_dv}. 
We show the results of the periodogram analyses in Figure \ref{fig:gls}. 
The periodogram analyzed for IRD RV measurements as well as for the activity indicators does not reveal any significant powers (FAP $<$ 5\%) for periods shorter than 60\,days, suggesting that ($i$) no RV signal from the transiting planet ($P=$ 12.76\,days) is detected (see Section\,\ref{sec:IRD_completeness} for completeness analysis), and ($ii$) the star is not chromospherically active.
The chromatic activity of \target was also investigated based on the measurement of the 
H$\alpha$ line by \citet{2017ApJ...834...85N}, who found no emission feature in the line core, with its EW being $\mathrm{EWH}\alpha = +0.064 \pm 0.024$ \r{A}; a positive sign of the EW value indicates that H$\alpha$ is an absorption, suggesting that \target is likely a Gyr-old star with low surface activity \citep[e.g.,][]{2021AJ....161..277K}. 
The insignificant H$\alpha$ emission is consistent with the weak X-ray activity (see Section \ref{sec:xray}) and the absence of flare-like events in the TESS light curves (Figure \ref{fig:lc_TESS}).
\target's X-ray luminosity ($L_{\rm X} = 6.0 \pm 1.3 \times 10^{25}$\,erg\,s$^{-1}$) is among the lowest detected in late-type stars \citep[e.g.,][]{car23} and indicates a quite low level of activity ($\log L_{\rm X}/L_{\rm bol} \approx -5.7$). \par

Finally, we searched for periodicity in archival light curves of \target to measure its rotation period ($P_{\rm{rot}}$).
We extracted 6 yr of light curves observed in the $g$ band between 2017 December and 2023 November from the ASAS-SN Sky Patrol\footnote{\url{https://asas-sn.osu.edu/}} \citep{2014ApJ...788...48S,2017PASP..129j4502K}.
We de-trended the ASAS-SN light curve using a quadratic function. 
We also analyzed the light curves from MEarth \citep[2008 October to 2020 February;][]{Berta_2012_MEarth} and SuperWASP \citep[2008 August to 2010 December;][]{Pollacco_2006}.
The MEarth light curve was obtained with Telescope 7.
The light curves were binned to nightly averages by first removing 3$\sigma$ outliers and then computing weighted averages of photometric measurements.
We performed GLS on them after 5$\sigma$ clipping, producing the power spectra in Figure \ref{fig:gls}. 
We found a significant periodicity at $\sim$85\,days in the power spectrum of ASAS-SN.
The power spectra of MEarth and SuperWASP show peaks at 83\,days and 90\,days, respectively, and these are close to $\sim$85\,days, though a higher peak appeared at $\sim$108\,days in the power spectrum of MEarth.  
Because the ASAS-SN light curve consists of the most continuous and frequently collected data points over many years among the three, we concluded that the periodicity at $\sim$85\,days is the most plausible value for $P_{\rm{rot}}$. 
We also measured $\log{R'_{\rm HK}}$ as in \citet{Perdelwitz_2021} on six public spectra taken with HARPS \citep{Mayor_2003} between 2006 and 2010.
The average value is in Table \ref{tab:stellar_para}.
Using the $\log{R'_{\rm HK}}$-$P_{\rm rot}$ relation of \citet{Astudillo-Defru_2017_RHK_Prot}, we estimated $P_{\rm rot}$ to be $P_{\rm rot,H\&K} = 60^{+22}_{-16}$\,days, which is consistent with the value measured in the ASAS-SN light curve. 
The same periodicity can also be seen in the periodogram for FWHM of IRD spectra at a FAP level of $\sim$5\%. We attributed these signals to spot modulation associated with the stellar rotation. This long rotational period also supports that \target is inactive and old. 

\begin{figure*}[]
\centering
\includegraphics[width=17.7cm,trim=0mm 0mm 0mm 0mm,clip]{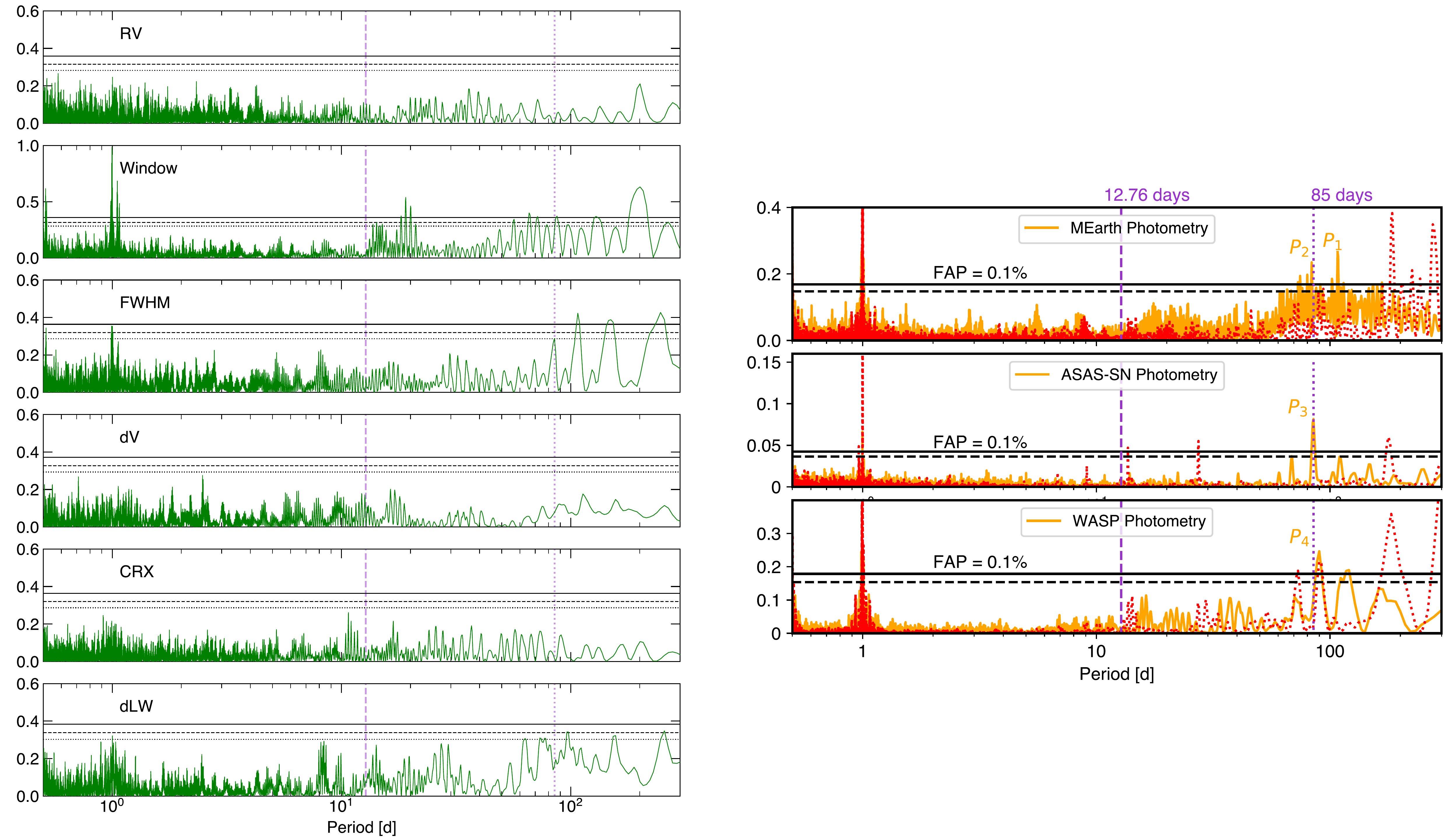}
\caption{
GLS periodogram analysis of the IRD RV measurements, the spectra, and the light curves of \target. 
The solid, dashed, and dotted horizontal lines in each panel correspond to FAPs of 0.1\%, 1\%, and 5\%, respectively.
The vertical dashed and dotted lines correspond to $P_{\rm orb} \approx$ 12.76\,days and $P_{\rm rot} \approx$ 85\,days, respectively.  
(Left) GLS powers (green lines) for RV, window function, FWHM, dV, CRX, and dLW are plotted in the panels from top to bottom.
(Right) GLS powers for the MEarth (top), ASAS-SN (middle), and SuperWASP (bottom) light curves.
Orange lines indicate GLS powers for the light curves, and red lines indicate their window functions. 
$P_{1}$ (= 108\,days), $P_{2}$ (= 83\,days), $P_{3}$ (= 85\,days), and $P_{4}$ (= 90\,days) in the panels indicate the periods of four peaks in the GLS powers.
} 
\label{fig:gls}
\end{figure*}

\subsection{Ruling Out False Positive Scenarios} \label{sec: FPScenario}

A grazing eclipsing binary (EB), a hierarchical EB (HEB), and a background EB (BEB) aligned with \target could mimic the transit signal detected in both TESS and MuSCAT series photometry. 
We ruled out those FP scenarios as follows.
The Gemini/NIRI $J$-band images observed in 2009 detected a $J \approx 17.9$\,mag source around \target, which was, however, undetected in the $K$-band images of NIRI (Section\,\ref{sec:imaging} and Appendix \ref{app:reduc_NIRI}).
We first assessed whether or not that faint source can be an FP source. 
The blue $J-K$ color ($<0.3$\,mag) of this source indicates that it must not be a companion to \target~but a distant early-type (F- or earlier-type) star \citep{Pecaut_2013_color}. 
Given the Gaia DR3 astrometry \citep{2023A&A...674A...1G}, this source is estimated to be separated from \target~by $\approx$ 9\farcs{0} in Sector\,43 of TESS, which is close enough to contaminate the TESS photometry. 
We constrained the source's brightness in the TESS band with the conversion functions from the TESS input catalog \citep[][their Equation \textit{jk} and \textit{joffset}]{2018AJ....156..102S} to be brighter than $T = 18.4 \pm 0.8$\,mag ($\Delta T < 8.1 \pm 0.8$\,mag).
Then, we converted the detected source's $J$-band magnitude to its $T$ magnitude under the limit of $J-K < 0.3$\,mag.
Thus, it could in principle be the cause of the transit signal (if a large fraction of its flux is occulted).
However, this source is located outside of most of the photometric apertures for the MuSCAT2 and MuSCAT3 observations (4\farcs1--7\farcs8 in radius for the first three transits), which finally confirmed the transit signal. 
This background source is therefore ruled out as the object causing the transit signal. 
Note that although the PDCSAP light curves do not correct for the dilution effect from this source due to the absence of this source in the Gaia\,DR3 catalog, the effect is so small that it can be neglected in the light-curve analyses (Sections\,\ref{sec:LC_TESS} and \ref{sec:joint}).  

Next, we considered other FP scenarios. 
Our RV data provide a mass constraint on the companion (see Section~\ref{sec:IRD_completeness}), ruling out the EB scenario. 
Furthermore, the constrained impact parameter ($b \lesssim 0.9$ at 3$\sigma$; see Section~\ref{sec:joint} and Table~\ref{tab:mcmc_results}) and boxy transit shape (see Figure~\ref{fig:joint-fit_lc}) eliminate the possibility of grazing transit geometries.
We simulated eclipses for a range of plausible HEBs following the method of \citet{2022Mori}, which was based on \citet{2020BoumaTOI837}.
Our simulations provided no configuration that can match the transit depth and transit shape observed in multiple bands with TESS and MuSCAT series.
The low value of the renormalised unit weight error (RUWE) from Gaia DR3 (1.198; see Table \ref{tab:stellar_para}) also supports that \target is a single star \citep{2020Belokurov}.

The probability of a chance alignment of a BEB is low given the target's location far from the Galactic plane ($b \approx -48\arcdeg$). 
Our AO imaging at $K^{\prime}$ rules out blends down to 0\farcs{2} for objects with $\Delta K^{\prime} < 5$\,mag (see Section \ref{sec:imaging}), and archival images reveal no bright background stars near the current position of \target (see Figure~\ref{fig:tess-aper}).

Finally, we computed the FP probability (FPP) of the case $P_{\rm{orb}}$ = 12.76\,days for \target with the Python software program {\tt TRICERATOPS} \citep{2020GiacaloneDressing}.  
We then used the {\tt TESS} light curve phase-folded at the given period, the Keck/NIRC2 AO contrast curve at $K^{\prime}$ given in Figure~\ref{fig:AOimage}, and the reported stellar parameters in Table~\ref{tab:stellar_para}. 
We ran {\tt TRICERATOPS} 20 times and computed an upper limit on the FPP of 0.7\% (3$\sigma$).
This value is below the threshold of FPP $< 1.5$\% prescribed by \citet{2021GiacaloneTOI}, further supporting that \target~b is a bona fide planet based on statistical validation.

\begin{figure*}[]
    \centering
    \includegraphics[width=18cm]{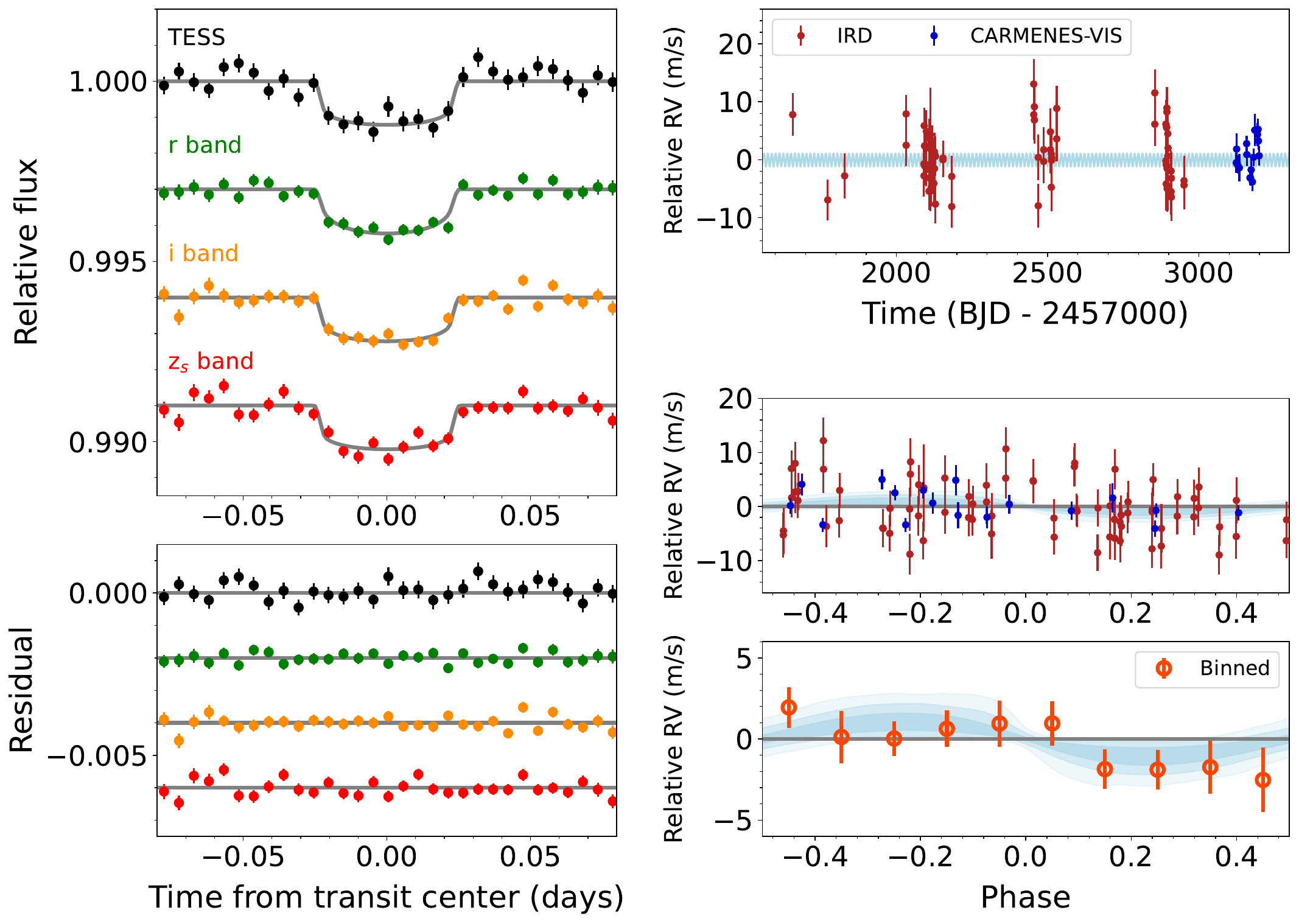}
    \caption{(Top left) Detrended (with the best-fit GP models), phase-folded, and 450 s binned transit light curves of \planet. The data taken in the TESS, $r$, $i$, and $z_s$ bands are plotted from top to bottom. The gray lines represent the best-fit transit models. Each light curve is vertically shifted by an arbitrary value for display purposes. (Bottom left) Same as the top left panel, but residuals from the best-fit transit models are shown.
    (Top right) RV time series of \target measured with IRD (red) and CARMENES (blue) along with the best-fit Keplerian model (light blue). 
    The error bars represent 1$\sigma$ uncertainties without including RV jitters. 
    (Middle right) Same as the right top panel, but {the data are} phase-folded with the orbital period of \target~b. The error bars here represent 1$\sigma$ uncertainties taking RV jitters into account. The darkest, second darkest, and lightest light-blue colors indicate the 1$\sigma$, 2$\sigma$, and 3$\sigma$ confidence regions calculated from the joint analysis, respectively. (Bottom right) Same as the right middle panel, but the data are binned into 10 equally spaced bins.}
    \label{fig:joint-fit_lc}
\end{figure*}
\begin{deluxetable*}{lccc}
\tablewidth{0pt}
\tablecaption{MCMC Fitting Results for Radial Velocity and Transit Light Curve Data\label{tab:mcmc_results}}
\tablehead{
        {\bf Parameter} & 
        {\bf Median and $\pm$1$\sigma$ interval} & 
        {\bf 3$\sigma$ range \tablenotemark{a}} &
        {\bf Prior \tablenotemark{b}}
}
\startdata
\multicolumn{3}{c}{Fitted parameters} \\ 
\hline
$b$ & $0.759\ ^{+0.049}_{-0.114}$ & (0.069, 0.904) & $\mathcal{U}$ (0, 1) \\ 
$k$ ($=R_{\rm{p}}/R_{\rm s}$) & $0.0336 \pm 0.0015$ & (0.0292, 0.0381) & $\mathcal{U}$ (0, 0.4) \\ 
$\sqrt{e}\sin{\omega}$ & $0.09 \pm 0.29$ & ($-$0.61, 0.66) & $\mathcal{U}$ ($-$1, 1) \\
$\sqrt{e}\cos{\omega}$ & $0.00 \pm 0.26$ & ($-$0.63, 0.65) & $\mathcal{U}$ ($-$1, 1) \\
$\ln \rho_{\rm s}$ ($\rho_\odot$) & $2.62 \pm 0.08$ & (2.38, 2.84) &  $\rho_{\rm s} = \mathcal{N}(13.5, 1.2)$ ($\rho_{\rm s} > 13.5$) \\
& & & $\rho_{\rm s} = \mathcal{N}(13.5, 0.9)$ ($\rho_{\rm s} \leq 13.5$)\\
$K$ (m s$^{-1}$) & $1.13\ ^{+0.56}_{-0.54}$ & $<$\,2.78 & $\mathcal{U}$ (0, $10^2$) \\
$\sigma_{\rm jit, IRD}$ (m s$^{-1}$) &  $2.80\ ^{+0.66}_{-0.69}$ & (0.20, 4.83)  & $\mathcal{U}$ (0, $10^2$) \\
$\sigma_{\rm jit, CAR}$ (m s$^{-1}$) &  $2.20\ ^{+0.76}_{-0.61}$ & (0.50, 5.37)  & $\mathcal{U}$ (0, $10^2$) \\
\hline
\multicolumn{3}{c}{Derived parameters}\\
\hline
Orbital period, $P$ (days) & $12.761408 \pm 0.000050$ & $\cdots$ & \\
Reference transit time, $T_{\rm{c,0}}$ (BJD) & $2459497.1865 \pm 0.0026$ & $\cdots$ & \\
Transit duration, $T_{14}$ (hr) & $1.20 \pm 0.09$ & (0.96, 1.74) & \\
Planet radius, $R_p$ ($R_{\oplus}$) & $0.958\ ^{+0.046}_{-0.048}$ & (0.815, 1.104)  & \\
Planet mass, $M_p$ ($M_{\oplus}$) & $1.57\ ^{+0.78}_{-0.75}$  & $<$\,3.87 & \\
Eccentricity, $e$ & $0.11\ ^{+0.14}_{-0.08}$ & $<$\,0.50  & \\
Orbital Inclination, $i_{\rm orb}$ (deg) & $89.194\ ^{+0.059}_{-0.052}$ & (89.013, 89.873)  & \\
Semimajor axis, $a$ (au) & $0.0668 \pm 0.0024$ & (0.0596, 0.0739)  & \\
Scaled semimajor axis, $a/R_s$ &  $54.9 \pm 1.4$ & (50.8, 59.3) & \\
Equilibrium temp., $T_{\rm eq}$ (K) (albedo=0) \tablenotemark{c} & $314.6\,^{+6.0}_{-5.4}$ & (299.5, 333.5)  & \\
Equilibrium temp., $T_{\rm eq}$ (K) (albedo=0.3) \tablenotemark{c}& $287.8\,^{+5.5}_{-5.0}$ & (274.0, 305.1)  & \\
Insolation, $S$ ($S_\oplus$) & $1.62\,^{+0.13}_{-0.11}$ & (1.33, 2.05) & \\
\enddata
\tablenotetext{a}{The two-sided 99.73\% confidence intervals are shown for all parameters except for the positive-valued parameters that are consistent with zero within 3$\sigma$ ($K$, $e$, and $M_p$), for which the one-sided 99.73\% confidence intervals are shown.}
\tablenotetext{b}{$\mathcal{U}(a, b)$ denotes a uniform distribution between $a$ and $b$, and $\mathcal{N}(a, b)$ denotes a normal distribution with the mean of $a$ and variance of $b^2$.}
\tablenotetext{c}{A uniform surface temperature is assumed.}
\label{tab:posteriors}
\end{deluxetable*}

\subsection{Joint Analysis of RVs and Transit Light Curves}
\label{sec:joint}

Although we did not detect any significant planetary signals in the RV data (Section \ref{sec:IRD_RV_Activity}), these data are still useful for placing upper limits on the planetary mass and orbital eccentricity. To best measure or constrain the orbit, mass, and radius of \planet, we modeled all the transit (TESS, MuSCAT2, and MuSCAT3) and RV (IRD and CARMENES) data simultaneously.
The model parameters for the RV data are as follows: RV semiamplitude $K$; orbital period $P$; two eccentricity parameters, $e\sin{\omega}$ and $e\cos{\omega}$, where $e$ and $\omega$ are the eccentricity and argument of periapsis, respectively; reference time of inferior (midtransit) passage $T_{\rm c,rv}$; RV zero-point for each instrument $v_{\rm 0, IRD}$ and $v_{\rm 0, CAR}$; and RV jitter for each instrument $\sigma_{\rm jit, IRD}$ and $\sigma_{\rm jit, CAR}$, where CAR stands for CARMENES.  

The parameters for the transit model are the same as for Sections \ref{sec:LC_TESS} and \ref{sec:LC_MuSCATs}, but this time the assumption of a circular orbit was removed. We also let individual midtransit times $T_{{\rm c}, i}$ for both the TESS and MuSCAT series data sets be free to take into account possible transit timing variations (TTVs). In addition, we used log stellar density $\ln \rho_s$ as a fitting parameter instead of $a_R$, which was converted from $\rho_s$ and $P$ using the equation $a_R = (\rho_s GP^2 / 3\pi)^{1/3}$ (assuming $k^3 \ll 1$), where $G$ is the gravitational constant.
We assumed that the radius ratio $k$ is common across all bands.

With the above parameterization, we performed an MCMC analysis using {\tt emcee}. In the analysis, we fixed $T_{\rm c, rv}$ and $P$ at the best-fit values of $T_{\rm c, 0}$ and $P$, respectively, derived from the analysis of the TESS data (Section \ref{sec:LC_TESS}). For the prior on $\ln \rho_{\rm s}$, we applied a two-sided Gaussian in the form of $\rho_s$ adopting the value estimated in Section \ref{sec:stellar_param} ($\rho_s = 13.5^{+1.2}_{-0.9}\,\rho_\odot$).
For the priors on the eccentricity parameters, we applied the joint probability distribution for $e$ and $\omega$ of \citet[][Equation (23) of that paper]{2014MNRAS.444.2263K}, who adopted the beta function as the underlying prior for $e$. For the two coefficients of the beta function, we adopted $\alpha=1.18$ and $\beta=6.34$ from \citet{Sagear_2023_eccentricity}, which were derived for singly transiting planets around early-to-mid-M dwarfs. 
The hyperparameters of the GP models for both the TESS and MuSCAT series' light curves were fixed to the best-fit values derived from the analyses in Sections \ref{sec:LC_TESS} and \ref{sec:LC_MuSCATs}, respectively.
For all the other parameters, uniform priors were applied. 
With a total of 28 free parameters, we ran 56 walkers $\times$ 10$^5$ MCMC steps after convergence for calculating posterior probability distributions. We assessed convergence by verifying that two independent MCMC runs yielded consistent posterior probability distributions for all parameters.

The median values and 1$\sigma$ boundaries of the fitting parameters and derived planetary parameters are reported in Table~\ref{tab:mcmc_results}. We found that \planet is an Earth-sized planet with a radius of $0.958^{+0.046}_{-0.048}\ \rear$.
We show detrended, phase-folded, and binned light curves from TESS, MuSCAT2, and MuSCAT3 created using the best-fit parameter values, as well as phase-folded RV data along with the posterior RV models, in Figure \ref{fig:joint-fit_lc}.
In Appendix~\ref{sec:posterior_of_joint-fit}, we provide a corner plot for the derived posterior probability distributions of a subset of the model parameters.

Applying a linear regression to the measured individual midtransit times $T_{\rm c, i}$, we refined $T_{\rm c, 0}$ and $P$ to be $2459497.1865 \pm 0.0026$ in BJD and $12.761408 \pm 0.000050$\,days, respectively.
In Appendix~\ref{sec:posterior_of_joint-fit}, we also show a plot of the midtransit times with the refined ephemeris subtracted. We found no significant TTVs in the data with a $\chi^2$ value of 8.6 with 8 degrees of freedom for the linear ephemeris fit.

\subsection{Completeness of IRD Data for Planet Detection} \label{sec:IRD_completeness}

We produced a detection completeness map of IRD RV measurements to constrain the presence of additional planets in the \target system.
To this end, we ran Monte Carlo simulations over the $\log{P_{\rm{orb}}}$ versus $\log{(m\sin{i})}$ grids, where $P_{\rm{orb}}$ and $m\sin{i}$ indicate the orbital period (in days) and the minimum mass (in units of $M_{\oplus}$) of a planet, respectively. 
We varied $\log{P_{\rm{orb}}}$ from 0 to 3 with 0.05 increments and $\log{(m\sin{i})}$ from 0 to 2 with 0.05 increments.
We randomly sampled the orbital parameters of a planet (eccentricity, argument of periapsis, and time of periastron) at each grid point, with \target's mass fixed to the value in Table \ref{tab:stellar_para}; the value of eccentricity was obtained from a Rayleigh distribution with $\sigma = 0.19$ \citep{Sagear_2023_eccentricity}, which is appropriate for a single planet around an early-to-mid-M dwarf. 
We then simulated a time series of Keplerians at each grid point, to which measurement uncertainties were added by sampling from zero-mean normal distributions with standard deviations equal to the internal errors of the IRD data set, as well as an additional RV jitter based on our estimate in Section \ref{sec:joint}. 
The periodogram analyses described in Section \ref{sec:IRD_RV_Activity} were applied to the artificial RVs, assuming a signal with power corresponding to FAP~$\leq$~1\% is a detection.
We repeated this procedure 100 times, thereby generating a map of detection statistics at each grid point with which to draw conclusions about the presence of planets around \target, which is shown in Figure \ref{fig:rv_detection_map}. \par

\begin{figure}[]
\centering
\includegraphics[width=\columnwidth, trim=0mm 0mm 15mm 10mm,clip]{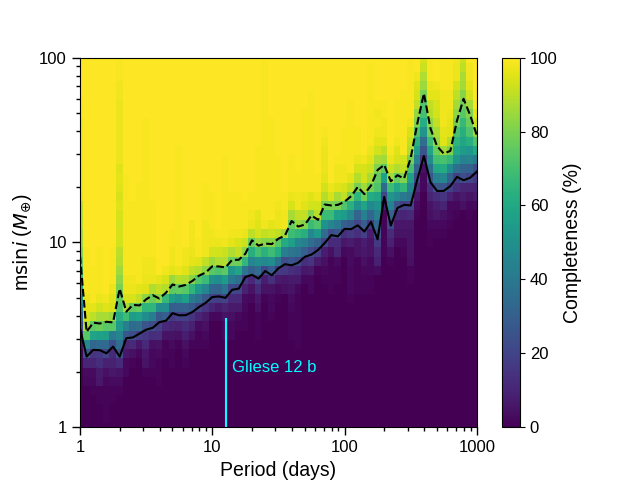}
\caption{Detection completeness of a simulated planet around \target. The horizontal axis and vertical axis indicate the orbital period and the minimum mass ($m\sin{i}$) of a planet, respectively. 
The color bar shows the percentage of instances of a simulated planet that is detected.
The solid and dashed lines correspond to the detection completeness of 20\% and 80\% of simulated planets, respectively.  
The vertical cyan line represents the orbital period ($\approx$ 12.76\,days) and mass range ($\leq$3$\sigma$) of \planet (see Section \ref{sec:joint}).}
\label{fig:rv_detection_map}
\end{figure}

Between 1 and 10 days, our simulations rule out the presence of a Neptune-mass planet (10--20\,$M_{\oplus}$) and show that there are most likely no massive super-Earths ($\approx$7--10\,$M_{\oplus}$).
Meanwhile, the existence of a low-mass super-Earth is not ruled out, as only 20\% of the total simulations were able to detect a planet as massive as $\sim$4\,$M_{\oplus}$ at $P_{\rm{orb}} =$ 5\,days.    
For $10$ days $< P_{\rm{orb}} < 30$\,days, the detection completeness is higher than 80\% only for planets heavier than massive super-Earths, while for planets with mass $\gtrsim$50 $M_{\oplus}$ it is above 80\% for almost the entire range of $P_{\rm{orb}}$.
In summary, our completeness simulations reveal that there are probably no massive super-Earths at $P_{\rm{orb}} = $1--30\,days, no Neptune-mass planets at $P_{\rm{orb}} <$ 150\,days, and no giant planets more massive than Saturn ($\approx$100 $M_{\oplus}$) at $P_{\rm{orb}} <$ 1000\,days.

\section{Discussion and Summary}\label{sec:Discussion}

We have identified an Earth-sized planetary candidate transiting the nearby mid-M dwarf \target from the TESS observations and subsequently validated its planetary nature from the ground-based follow-up observations including multiband transit photometry with MuSCAT2 and MuSCAT3, high-contrast imaging with NIRI and NIRC2, and RV measurements with IRD and CARMENES. We have revealed that the planet, \planet, has a radius of $0.96 \pm 0.05\,R_\oplus$, a $3\sigma$ mass upper limit of $3.9\,M_\oplus$, and an orbital period of 12.76\,days.
We have found no additional planet with a mass larger than Neptune (Saturn) within an orbital period of 50\,days (1000\,days) based on a 4 yr long RV monitoring with IRD.

Figure \ref{fig:MR} shows the mass-radius relation for planets less massive than $6\,M_\oplus$ around M dwarfs.
The weak constraint on the mass of \planet from our RV measurements allows various internal compositions for this planet.
Within the measured upper limit on the mass, \planet could be a dense Earth-sized planet (even as dense as pure iron).
This possibility is intriguing because such a high-density planet has only rarely been discovered around relatively metal-poor stars \citep[e.g.,][]{2022A&A...665A.154B} such as \target ([Fe/H]$=-0.32 \pm 0.06$).
Alternatively, \planet could be volatile-rich. 
If the mass of \planet is well below 1$\mear$, then \planet could be either a rocky planet with a small mass-fraction of a hydrogen-rich atmosphere or a water-rich planet without a hydrogen-rich atmosphere (see Figure \ref{fig:MR}), the latter of which would indicate either orbital migration or an efficient inward transport of icy material from outside the snow line.
Thus, precisely determining the mass of \planet via follow-up RV observations is important in the context of the formation and evolution of Earth-sized planets and will undoubtedly be the subject of future works. \par

\begin{figure*}[]
    \centering
    \includegraphics[scale=0.72,trim=10mm 0mm 0mm 0mm,clip]{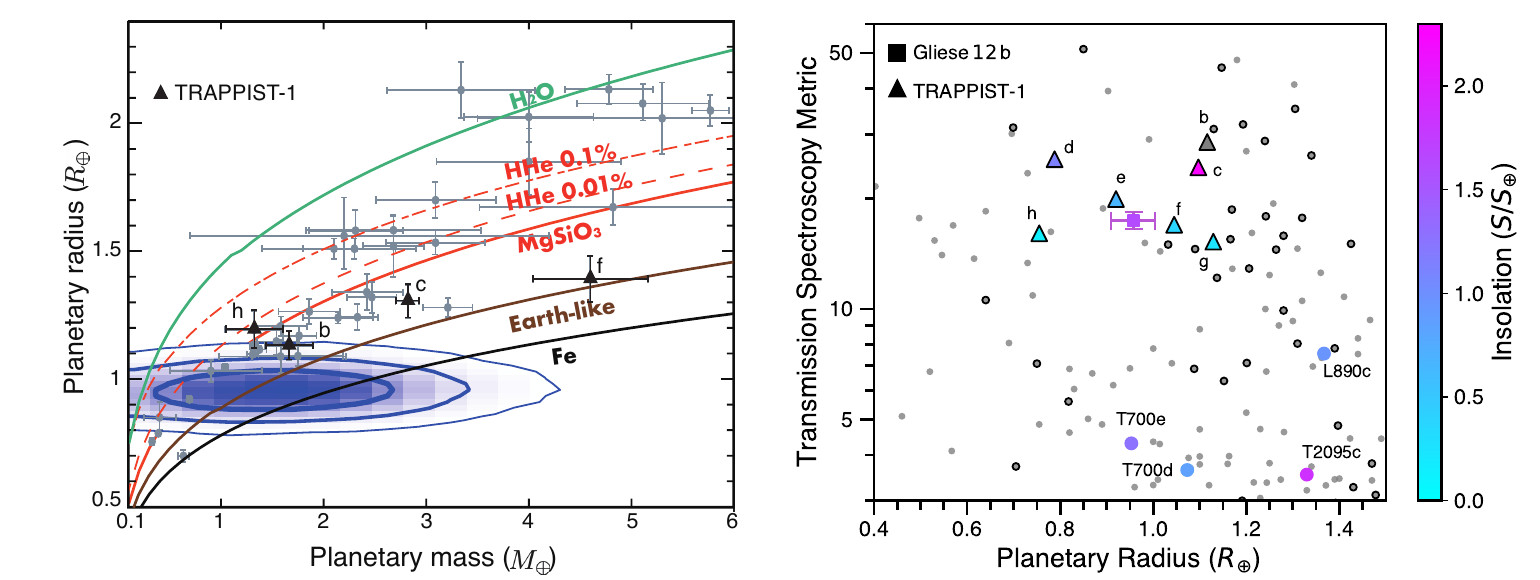}
    \caption{ (Left) Mass-radius relation for planets with masses below $6 M_\oplus$ around M dwarfs with $T_\mathrm{eff} \leq 3900$\,K. Gray circles are all the known exoplanets in this mass range taken from \citet{NASA_Exoplanet_mr}. We plot the 1$\sigma$, 2$\sigma$, and 3$\sigma$ contours of the mass and the radius of \planet.  The colored curves show interior models of planets that are composed of pure water, pure silicate, Earth-like rock, and pure iron. We adopted $T_\mathrm{eq} = 312.9$\,K and AQUA \citep{2020A&A...643A.105H}, Birch-Murnaghan, and Vinet EoS \citep{2007ApJ...669.1279S, 2013PASP..125..227Z} as equations of state of H$_2$O, MgSiO$_3$, and $\epsilon-$Fe.
    The dashed and dashed-dotted curves show the mass-radius relation of a rocky core with a hydrogen-rich atmosphere of 0.01\,wt\% and 0.1\,wt\%, respectively.
    (Right) TSM values calculated for \planet (square) and other small ($R \leq 1.5\rear$) exoplanets against their planetary radii.      
    Colored symbols indicate the planets whose insolation flux ($S$) is $< 2.3$ $S_{\oplus}$, with colors scaled by $S$ (as shown by the color bar). 
    The symbols with black edges represent the planets with masses constrained by methods other than an empirical mass-radius relation. 
    The planets within this range are labeled: T700, L890, and T2095 stand for TOI-700 \citep{Gilbert_2023_TOI700}, LP\,890-9 \citep{Delrez_2022}, and TOI-2095 \citep{Murgas_2023_TOI2095}, respectively.
    The planets with $S > 2.3\ S_\oplus$ are indicated by gray symbols.
    On both the left and right panels, the planets around TRAPPIST-1 are plotted with triangles.    
    }
    \label{fig:MR}
\end{figure*}

With an orbital period of $~12.76$\,days, \planet is likely to be tidally locked. 
The insolation flux that \planet receives from the host star is calculated to be 1.62$^{+0.13}_{-0.11}$ $S_\oplus$, where $S_\oplus$ is the insolation of the Earth. 
This makes its location even closer to the star than the inner edge of the HZ for a tidally locked planet orbiting a star with an effective temperature of 3300\,K \citep[$\approx$1.3\,$S_\oplus$,][]{Kopparapu_2017}. 
Thus, \planet is less likely to retain a stable ocean, although the inner edge of the HZ could depend on several other factors as well, such as the total atmospheric pressure, the planetary mass, and the land-to-ocean fraction \citep[e.g.,][]{Kodama2019}. 
\planet is more likely to be in the runaway greenhouse state, where the planetary atmosphere contains water vapor, or in a dry state after water loss by escape, where the atmosphere contains little water vapor.
\planet is therefore an intriguing target for future atmospheric studies to understand the atmospheric state of potentially rocky planets near the inner edge of the HZ.

The proximity ($d = 12$\,pc) and brightness ($J = 8.6$\,mag) of the system make \planet a still uncommon, low-insolation, and potentially terrestrial planet that is amenable to atmospheric characterization through transmission spectroscopy.
Using the empirical mass-radius relationships of \cite{Chen_2017_mass_radi} and \cite{Louie_2018},  we convert \planet's radius to a mass of 0.83$^{+0.15}_{-0.14}$ $\mear$. 
With this mass determination, we followed \citet{Kempton_2018_TSM} to determine a transmission spectroscopy metric (TSM) value of \planet to be 17.4$^{+1.0}_{-0.9}$, which is compared with known planets with radii smaller than 1.5 $\rear$ in the right panel of Figure\,\ref{fig:MR}.
We note that the theoretical mass-radius relation shown in Figure \ref{fig:MR} limits the TSM value of \planet to be higher than $\sim$7.
The higher the TSM value of a planet, the more suitable it is for atmosphere transmission spectroscopy \citep[see][]{Kempton_2018_TSM}.
All the data for the known (confirmed) planets are taken from \citet{NASA_Exoplanet_tsm}\footnote{For those planets lacking a proper mass constraint in the catalog (93 out of 133), we estimated the mass using the radius and a mass-radius relation in the same way as was done for \planet.}.
In the figure, we highlight the planets that receive insolation fluxes ($S$) lower than 2.3 $S_{\oplus}$ (which is approximately equivalent to the insolation flux onto TRAPPIST-1\,c).  
Within the parameter range shown here, only the TRAPPIST-1 planets and \planet have TSM values higher than 10, placing \planet in the still limited sample of temperate and potentially terrestrial planets that are suitable for transmission spectroscopy.   

The actual detectability of atmospheric signatures, however, depends on the amount and composition of the atmosphere, which can be affected by the XUV irradiation from the host star.
A recent JWST/MIRI observation revealed that TRAPPIST-1 c has no thick atmosphere like that of Venus, which may be due to atmospheric escape caused by strong XUV radiation and stellar winds from the host star \citep{Zieba_2023_TRAPPIST1c,Lincowski_2023,Teixeira_2024_TRAPPSIT1c}. 
The semimajor axis of TRAPPIST-1\,c is 0.016\,au \citep{Agol_2021_TRAPPIST1} from the host, which has an XUV luminosity of $L_{\rm XUV} \approx 0.7$--$1.8 \times 10^{27}$\,erg\,s$^{-1}$\citep{2017MNRAS.465L..74W,2020AJ....159..275B}. 
On the other hand, we used our coronal model as in \citet[][and references therein]{san11} to estimate the XUV luminosity of \target to be $L_{\rm XUV} \approx 4.4 \times 10^{26}$\,erg\,s$^{-1}$ (for the range of 5--920\,\AA), which is even lower than that of TRAPPIST-1.
Since the semimajor axis of \planet (0.067\,au) is nearly 4 times larger than that of TRAPPIST-1\,c, \planet is approximately 30--70 times less irradiated in XUV than TRAPPIST-1\,c.
Even with this currently weak XUV irradiation, \planet may have lost a large part of its  primordial (hydrogen-rich) atmosphere considering that the XUV luminosity of the star must have been higher in the past \citep[e.g.,][]{Penz_2008_XrayMdwaf,Johnstone_2021_XrayEvo}. Nevertheless, there is a higher probability that \planet has a more significant secondary (C-, N-, and/or O-rich) atmosphere than TRAPPIST-1\,c, 
because the amount of secondary atmosphere is considered to be regulated by the supply of gases (e.g., outgassing) and the escape driven by the current XUV irradiation and stellar wind. 

Atmospheric characterization of \planet using transmission spectroscopy with JWST would therefore allow direct comparison to other Earth-sized exoplanets like the TRAPPIST-1 worlds.  
To explore the feasibility of observing the atmosphere of \planet with JWST, we simulate its transmission spectrum using the radiative transfer and retrieval code \texttt{TauREx3} \citep{Al-Refaie_2021}, assuming that the density of \planet is comparable to that of the Earth. Since the actual atmospheric composition of this planet is difficult to know {\it a priori}, we explore five simple cases to highlight the telescope performances:

\begin{itemize}
    \item Case 1: a reference scenario with remaining primordial gas (H$_2$ and He). The atmosphere is simulated at chemical equilibrium using the {\tt GGChem} code \citep{Woitke_2021} with a metallicity of 100 $\times$ solar.
    \item Case 2: a Titan-like atmosphere composed of 95\% N$_2$ and 5\% CH$_4$ with constant altitude abundance profiles. A fully opaque cloud layer at 0.1 bar is also added.
    \item Case 3: a water-world scenario with a 100\% H$_2$O atmosphere and a fully opaque cloud layer similar to case 2.
    \item Case 4: a Venus-like case with a 100\% CO$_2$ atmosphere and a fully opaque cloud layer similar to case~2. 
    \item Case 5: a clear Venus-like case. Same as case 4 but without clouds.
\end{itemize}

In these models, the atmosphere is simulated using the \texttt{TauREx} transit forward model and assuming a plane-parallel atmosphere with 100 layers evenly spaced in log-space between 10\,bar and $10^{-5}$\,bar. 
We assumed an isothermal temperature-pressure profile at the equilibrium temperature of the planet. We modeled the atmosphere at the native resolution $\mathcal{R} = 15,000$ of the cross-section (H$_2$O: \citealt{Polyansky_2018}; CH$_4$: \citealt{Yurchenko_2017,Chubb_2020}; CO$_2$: \citealt{Chubb_2020,Yurchenko_2020}) and then convolved the resulting spectrum with the JWST instrument response function for NIRSpec \citep{2022A&A...661A..80J} grisms G140H, G235H, and G395H, which we obtained via the online {\tt Pandexo} tool \citep{Batalha_2017}. We note that to avoid $>$80\% saturation, G140H and G235H require three and five groups, respectively, which would make the characterization of the nondestructive ramps more difficult than with G395H, for which seven groups is optimal. We display the final models and observed spectra for six stacked observations with each JWST instrument configuration in Figure~\ref{fig:spectra}. Note that the simulated observations have been binned downed to $\mathcal{R} = 50$ for visualization purposes. The simulated JWST transmission spectra of \target\,b demonstrate detectable atmospheric features for a range of compositions, underscoring the potential to characterize this terrestrial exoplanet using a modest amount of JWST time. 
Furthermore, transit photometry with JWST for an active star like TRAPPIST-1 is disturbed by the contamination of flares \citep{Howard_2023}, which should be less problematic for \target due to its quiescent nature.\par
Given that \target is only the second M dwarf system within 30\,pc hosting Earth-sized planets with insolation less than 2\,$S_\oplus$, following TRAPPIST-1, it will undoubtedly be the subject of many future studies as a benchmark system. These studies will help reveal the diversity of atmospheres and climates on temperate rocky planets, furthering our understanding of planetary environments beyond our solar system.

\begin{figure*}[]
   \centering
   \includegraphics[width=\textwidth]{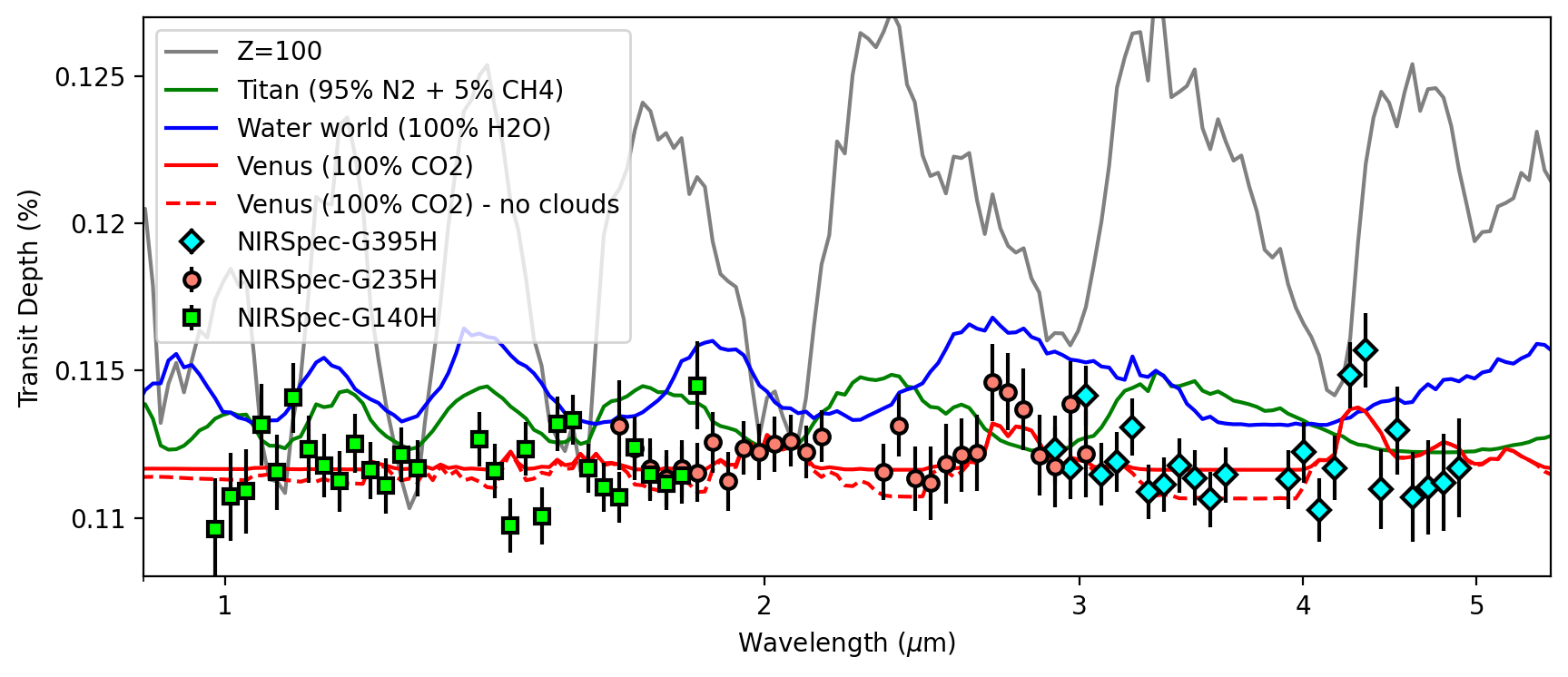}
   \caption{Predicted transmission spectra for \target\,b, assuming an isothermal temperature-pressure profile, an Earth-like density, and several different cloud-free compositions: primordial H$_2$ and He at 100$\times$solar metallicity, Titan-like, water world, and Venus-like. For the Venus-like scenario, we also show a model including 0.1 bar fully opaque clouds. We show simulated data assuming the cloudy Venus scenario and six transits per JWST/NIRSpec grism: G140H (green squares), G235H (red circles), and G395H (blue diamonds).}
   \label{fig:spectra}
\end{figure*}

\section*{Acknowledgments}

We thank the reviewer, whose careful reading of the manuscript and excellent suggestions have improved the quality of this paper.
We thank Andrew W. Stephens and Sandy Leggett for data reduction support for the Gemini/NIRI data. We acknowledge that the TRES team posted the results of their spectroscopic observations of \target spanning 5 yr to ExoFOP-TESS soon after TOI-6251.01 was released, showing no large RV variations  (RMS $=$ 21\,m\,s$^{-1}$).

Funding for the TESS mission is provided by NASA's Science Mission Directorate. We acknowledge the use of public TESS data from pipelines at the TESS Science Office and at the TESS Science Processing Operations Center (SPOC). Resources supporting this work were provided by the NASA High-End Computing (HEC) Program through the NASA Advanced Supercomputing (NAS) Division at Ames Research Center for the production of the SPOC data products. This paper includes data collected by the TESS mission that are publicly available from the Mikulski Archive for Space Telescopes (MAST).
All the TESS data used in this paper can be found in MAST: \dataset[https://doi.org/10.17909/psfb-yg51]{https://doi.org/10.17909/psfb-yg51}.

The Gemini data that we used in this work are publicly available from the Gemini Observatory Archive (GOA; \url{https://archive.gemini.edu}) under the program GN-2009B-Q-10 (PI: Sergio Dieterich), whose abstract is shown at \url{https://archive.gemini.edu/programinfo/GN-2009B-Q-10}. 
This work is based on observations obtained with XMM-Newton, an ESA science mission
with instruments and contributions directly funded by ESA Member States and NASA.
This work has made use of data from the European Space Agency (ESA) mission Gaia \citep[][\url{https://www.cosmos.esa.int/gaia}]{2016A&A...595A...1G}, processed by the Gaia Data Processing and Analysis Consortium (DPAC, \url{https://www.cosmos.esa.int/web/gaia/dpac/consortium}). Funding for the DPAC has been provided by national institutions, in particular the institutions participating in the Gaia Multilateral Agreement.
This work used the data from the Second Palomar Observatory Sky Survey (POSS-II), which was made by the California Institute of Technology with funds from the National Science Foundation, the National Geographic Society, the Sloan Foundation, the Samuel Oschin Foundation, and the Eastman Kodak Corporation.
This paper makes use of data from the MEarth Project, which is a collaboration between Harvard University and the Smithsonian Astrophysical Observatory. The MEarth Project acknowledges funding from the David and Lucile Packard Fellowship for Science and Engineering, the National Science Foundation under grants AST-0807690, AST-1109468, AST-1616624 and AST-1004488 (Alan T. Waterman Award), the National Aeronautics and Space Administration under grant No. 80NSSC18K0476 issued through the XRP Program, and the John Templeton Foundation.
This work is based on HARPS data obtained from the ESO Science Archive Facility: the HARPS data are based on observations made with ESO Telescopes at the La Silla Paranal Observatory under program ID 072.C-0488. 
This research has made use of the SIMBAD and VizieR services, both operated at Centre de Donn{\' e}es astronomiques de Strasbourg (CDS, \url{https://cds.u-strasbg.fr/}) in France, and NASA's Astrophysics Data System Bibliographic Services.
This research made use of Lightkurve, a Python package for Kepler and TESS data analysis \citep{Lightkurve_2018}.
We used \texttt{TESSCut} \citep{TESScut} in the analysis that was based on \texttt{TRICERATOPS} \citep{2020GiacaloneDressing}.
This publication makes use of data products from the Two Micron All Sky Survey \citep{doi-2MASS}, which is a joint project of the University of Massachusetts and the Infrared Processing and Analysis Center/California Institute of Technology, funded by the National Aeronautics and Space Administration and the National Science Foundation.
This research has made use of the NASA Exoplanet Archive (\doi{10.26133/NEA12}), which is operated by the California Institute of Technology, under contract with the National Aeronautics and Space Administration under the Exoplanet Exploration Program.
IRAF, which we used in reducing IRD raw data, is distributed by the National Optical Astronomy Observatories, which is operated by the Association of Universities for Research in Astronomy, Inc. (AURA) under cooperative agreement with the National Science Foundation.
This research has made use of the Exoplanet Follow-up Observation Program (ExoFOP; DOI: \doi{10.26134/ExoFOP5}) website, which is operated by the California Institute of Technology, under contract with the National Aeronautics and Space Administration under the Exoplanet Exploration Program.
Part of the data analysis was carried out on the Multi-wavelength Data Analysis System operated by the Astronomy Data Center (ADC), National Astronomical Observatory of Japan. 
Part of this work was carried out at the Jet Propulsion Laboratory, California Institute of Technology, under contract with NASA.\par

We are honored and grateful for the opportunity of observing the Universe from Maunakea, which has cultural, historical, and natural significance in Hawaii.
We appreciate the critical support from all the current and recent Subaru and Keck Observatory staffs. 
Their support was essential in achieving this discovery, especially amidst the many difficulties associated with the COVID-19 pandemic. \par

This research is based on data collected at the Subaru Telescope, which is operated by the National Astronomical Observatory of Japan. 
This article is based on observations made with the MuSCAT2 instrument, developed by ABC, at Telescopio Carlos Sánchez operated on the island of Tenerife by the IAC in the Spanish Observatorio del Teide.
This paper is based on observations made with the MuSCAT3 instrument, developed by the Astrobiology Center and under financial support by JSPS KAKENHI (JP18H05439) and JST PRESTO (JPMJPR1775), at Faulkes Telescope North on Maui, Hawaii, operated by the Las Cumbres Observatory.
CARMENES is an instrument at the Centro Astron\'omico Hispano en Andaluc\'ia (CAHA) at Calar Alto (Almer\'{\i}a, Spain), operated jointly by the Junta de Andaluc\'ia and the Instituto de Astrof\'isica de Andaluc\'ia (CSIC).
Some of the data presented herein were obtained at the W. M. Keck Observatory, which is operated as a scientific partnership among the California Institute of Technology, the University of California, and the National Aeronautics and Space Administration. The Observatory was made possible by the generous financial support of the W. M. Keck Foundation.

M.T. is supported by JSPS KAKENHI grant Nos. 18H05442, 15H02063, and 22000005.
This work is partly supported by JSPS KAKENHI grant Nos JP18H05439, JP21K13955, JP21K20376 and JST CREST grant Nos JPMJCR1761. 
M.O. is funded by the National Natural Science Foundation of China (Nos. 12250610186, 12273023).
We acknowledge financial support from the Agencia Estatal de Investigaci\'on (AEI/10.13039/501100011033) of the Ministerio de Ciencia e Innovaci\'on and the ERDF ``A way of making Europe'' through projects 
  PID2021-125627OB-C31,		
    PID2021-125627OB-C32,		
  PID2019-109522GB-C5[1:4],	
the grant PRE2020-093107 of the Pre-Doc Program for the Training of Doctors (FPI-SO) through FSE funds, 
and the Centre of Excellence ``Severo Ochoa'' and ``Mar\'ia de Maeztu'' awards to the Instituto de Astrof\'isica de Canarias (CEX2019-000920-S) and Institut de Ci\`encies de l'Espai (CEX2020-001058-M).
K.A.C acknowledges support from the TESS mission via subaward s3449 from MIT.
D.R. was supported by NASA under award No. NNA16BD14C for NASA Academic Mission Services.
R.L. acknowledges funding from the University of La Laguna through the Margarita Salas Fellowship from the Spanish Ministry of Universities ref. UNI/551/2021-May 26 and under the EU Next Generation funds. 
M.S. acknowledges the support of the Italian National Institute of Astrophysics (INAF) through the project 'The HOT-ATMOS Project: characterizing the atmospheres of hot giant planets as a key to understand the exoplanet diversity' (1.05.01.85.04).
G.M. acknowledges fundings from the Ariel Postdoctoral Fellowship program of the Swedish National Space Agency (SNSA).
J.K. gratefully acknowledges the support of the Swedish Research Council  (VR: Etableringsbidrag 2017-04945).
We acknowledge the support from the Deutsche Forschungsgemeinschaft (DFG) under Research Unit FOR2544 ``Blue Planets around Red Stars'' through project DR~281/32-1.
The results reported herein benefited from collaborations and/or information exchange within NASA’s Nexus for Exoplanet System Science (NExSS) research coordination network sponsored by NASA’s Science Mission Directorate under agreement No. 80NSSC21K0593 for the program ``Alien Earths".
Part of this work was also supported by NAOJ Research Coordination Committee, NINS, NAOJ-RCC2301-0401.
\\
\facility{TESS, Keck:II (NIRC2), Gemini-North (Altair, NIRI), the 1.52\,m Telescopio Carlos S\'anchez (MuSCAT2), the 2 m Faulkes Telescope North (MuSCAT3), the Subaru Telescope (IRD), the 3.5 m Calar Alto telescope (CARMENES), XMN-Newton, the ESO La Silla 3.6m telescope (HARPS), SuperWASP, ASAS-SN, MEarth, Exoplanet Archive}
\software{\texttt{astropy}: \citet{astropy_2013,astropy_2018,astropy_2022}, {\tt PyAstronomy}: \citet{pyastronomy}; \url{https://github.com/sczesla/PyAstronomy}, {\tt corner}: \cite{corner}, {\tt celerite}: \citet{2017AJ....154..220F}, {\tt emcee}: \citet{2013PASP..125..306F}, {\tt GGChem}: \citet{Woitke_2021}, {\tt IRAF}: \citet{Tody_IRAF_1986, Tody_1993_IRAF}, {\tt IRD RV-measurement pipeline}: \citet{Hirano_2020_IRDPipeline}, {\tt lighkurve}: \citet{Lightkurve_2018}, {\tt LDTk}: \citet{Parviainen2015b}, \texttt{matplotlib}: \citet{Hunter_2007}, \texttt{nirlin}: \url{https://www.gemini.edu/instrumentation/niri/data-reduction}, {\tt numpy}: \citet{van_der_Walt_2011_numpy}, {\tt Pandexo}: \citet{Batalha_2017}, {\tt PyTransit}: \citet{Parviainen2015}, \texttt{scipy}: \citet{Virtanen_2020_scipy}, \texttt{serval}: \citet{Zechmeister_2018_SERVAL}, {\tt SteParSyn}: \citet{Tabernero_2022}, \texttt{TauREx3}: \citet{Al-Refaie_2021}, TESSCut: \citet{TESScut}, {\tt TRICERATOPS}: \citet{2020GiacaloneDressing}, \texttt{pandas}: \citet{Pandas}, \texttt{SymPy}: \cite{SymPy_2017}, Google Colaboratory}

\appendix

\section{High Contrast Imaging and Data Reduction} \label{app: highcontrast_detail}
 
\subsection{Gemini/NIRI}\label{app:reduc_NIRI} 

The near-infrared observations of \target were performed with the f/32 camera of NIRI and the Altair field lens, which set the plate scale to be 21.4 mas pixel$^{-1}$. 
A partial readout mode was employed to image only the central 512 $\times$ 512 pixels of the full FoV of the Altair+NIRI combination. 
The $J$-, $H$-, and $K$-band observations obtained six images with dithering. 
At each dither position, a single frame was made by integrating five subframes (i.e., five coadds) taken with individual exposures of 0.15 s in $H$- and $K$-band imaging and 1.2 s in $J$-band imaging.  \par

We reduced the archival Gemini/NIRI data with a custom-made script as follows. 
In the script, we also used a Python program for linearity correction developed by the NIRI instrument team, which is \texttt{nirlin.py} and available from \url{https://www.gemini.edu/instrumentation/niri/data-reduction}.
We performed the raw-data processing in this order: linearity correction, horizontal-stripe pattern subtraction, flat-fielding, sky subtraction, bad-pixel interpolation, distortion correction, and alignment of \target's PSF centers.
The horizontal-stripe pattern of the detector was removed by subtracting each row of the detector by its median flux after masking out pixels containing flux from the star.
Based on communication with NIRI instrument expert A. Stevens, we did not apply \texttt{cleanir.py} for subtracting the eight-column noise pattern, since the pattern was not present in the images. 
The script for linearity correction cannot handle data with image sizes of 512 $\times$ 512 pixels. 
We therefore applied all correction functions that are appropriate for other image sizes to the data and tested whether the output contrast measurements settled into consistent results regardless of the adopted functions.
A total of 99\% of the measurements are consistent within 0.05 mag, indicating that there are no significant differences between different nonlinearity corrections.
The images were taken with a fixed Cassegrain rotator in order to optimize the AO corrections. 
In this mode, the field rotates; we therefore derotated the images based on the information in the fits headers to align the FoVs to the north direction.    
Finally, we combined all the processed images at $J$, $H$, and $K$ into three master images at each band (see Figure \ref{fig:AOimage}). 

\subsection{Keck II/NIRC2+PyWFS}\label{app:reduc_NIRC2} 

We obtained two cycles of three $K^{\prime}$-band ($\approx$2.12 $\mu$m) images at different dithering positions. 
Then, each image of the first cycle was obtained by coadding five subimages taken with 1 s exposure, while 10 subimages of 1 s exposure were coadded at each dithering position in the second cycle.   We also obtained six images with a narrowband filter corresponding to the Br$\gamma$ feature ($\approx$2.17$\mu$m) for aperture corrections.
The Br$\gamma$ images were used as the unsaturated reference PSFs to compensate the saturated cores of PSFs in the photometry of the $K^{\prime}$-band images under the assumption that the PSF shapes are quite similar between $K^{\prime}$ and Br$\mathrm{\gamma}$. 
We checked whether the assumption affected the contrast measurements with the unsaturated PSFs of the other two M dwarfs observed in the same night. 
The measurements are in agreement within 10\%, ensuring that the above assumption does not influence our conclusions. \par
We performed dark subtraction, flat-fielding correction, and bad-pixel removal to calibrate the raw NIRC2 data.  
We next calibrated the camera's optical distortion with the script provided by \citet{Yelda_2010} and the distortion solution from \citet{Service_2016_NIRC2}, which sets the plate scale to be 9.971 mas pixel$^{-1}$ in the distortion-corrected image.
We combined the calibrated images after shifting the primary star's PSFs to a common position.
Before combining the images, we also rotated them to align in the north direction along the $y$-axis of the images because we conducted the observations with the pupil-tracking mode. 
Figure \ref{fig:AOimage} shows the combined image. \par

\section{RV Measurements of \target } \label{sec: RV_measures}
Table \ref{tab:rv_GJ12_IRD} lists \target's relative RVs measured with IRD and CARMENES. 
\startlongtable
\begin{deluxetable}{lcll}\label{tab:rv_GJ12_IRD}
     \tablewidth{0pt}
    \tablecaption{Relative RV Measurements of \target}
    \tablehead{\colhead{BJD (days)} & \colhead{RV (m s$^{-1}$)} &  \colhead{$\sigma_{\rm{RV}}$ (m s$^{-1}$)} & \colhead{Instrument}}
    \startdata
2458657.1081023 & 7.78 & 3.68 & IRD\\
2458772.8545807 & $-$6.94 & 3.57 & IRD\\
2458828.7733195 & $-$2.78 & 3.89 & IRD\\
2459032.1058368 & 7.91 & 3.31 & IRD\\
2459032.1108176 & 2.50 & 3.57 & IRD\\
2459091.1383862 & $-$0.69 & 3.61 & IRD\\
2459091.1444178 & $-$2.78 & 3.61 & IRD\\
2459091.9128072 & 5.87 & 3.07 & IRD\\
2459092.8994228 & $-$1.04 & 3.31 & IRD\\
2459092.9072207 & 2.39 & 3.49 & IRD\\
2459098.9899503 & $-$0.85 & 3.69 & IRD\\
2459098.9954901 & 4.91 & 3.67 & IRD\\
2459107.8986615 & $-$5.44 & 3.24 & IRD\\
2459107.9064719 & $-$1.58 & 3.37 & IRD\\
2459108.8313943 & 1.98 & 3.26 & IRD\\
2459108.8421331 & 3.68 & 3.41 & IRD\\
2459109.8320718 & $-$1.73 & 3.09 & IRD\\
2459109.8428024 & 3.87 & 3.23 & IRD\\
2459110.8857408 & $-$3.09 & 3.56 & IRD\\
2459110.8964816 & $-$3.19 & 3.31 & IRD\\
2459111.8681702 & $-$5.44 & 3.48 & IRD\\
2459111.8812654 & 4.34 & 8.06 & IRD\\
2459112.9655217 & $-$1.16 & 3.43 & IRD\\
2459112.9744962 & 2.75 & 3.18 & IRD\\
2459115.0326667 & $-$1.26 & 3.51 & IRD\\
2459115.0387004 & $-$4.78 & 3.22 & IRD\\
2459118.0189589 & $-$0.90 & 3.43 & IRD\\
2459118.0261636 & 2.70 & 3.27 & IRD\\
2459123.8158675 & $-$4.08 & 3.36 & IRD\\
2459123.8236605 & 0.53 & 3.31 & IRD\\
2459125.8172895 & $-$1.52 & 3.00 & IRD\\
2459125.8280198 & 1.36 & 3.34 & IRD\\
2459128.8497845 & $-$7.66 & 3.35 & IRD\\
2459128.8605252 & 0.60 & 3.44 & IRD\\
2459153.8646901 & 0.28 & 2.98 & IRD\\
2459153.8765642 & 0.00 & 2.98 & IRD\\
2459182.8403137 & $-$8.08 & 3.68 & IRD\\
2459182.8510060 & $-$2.88 & 3.55 & IRD\\
2459454.0011469 & 13.07 & 4.28 & IRD\\
2459454.0048046 & 7.78 & 4.39 & IRD\\
2459456.1086614 & 6.86 & 4.11 & IRD\\
2459456.1140850 & 9.15 & 4.34 & IRD\\
2459468.8528721 & 0.41 & 3.98 & IRD\\
2459468.8573561 & $-$7.94 & 3.82 & IRD\\
2459486.8983724 & $-$0.30 & 3.78 & IRD\\
2459486.9020344 & 1.73 & 3.90 & IRD\\
2459508.9949030 & 4.82 & 4.05 & IRD\\
2459508.9985653 & 1.73 & 4.05 & IRD\\
2459511.9844461 & $-$4.75 & 4.18 & IRD\\
2459511.9892896 & 1.00 & 4.11 & IRD\\
2459513.0088874 & $-$0.08 & 3.61 & IRD\\
2459513.0149102 & 0.34 & 4.01 & IRD\\
2459529.8861489 & 8.85 & 3.94 & IRD\\
2459529.8900382 & 3.60 & 3.90 & IRD\\
2459854.0161688 & 6.13 & 3.82 & IRD\\
2459854.0210022 & 11.55 & 4.01 & IRD\\
2459890.8249876 & 6.15 & 4.22 & IRD\\
2459890.8286465 & $-$0.19 & 3.93 & IRD\\
2459891.9583209 & $-$4.19 & 4.55 & IRD\\
2459891.9631482 & $-$0.87 & 4.26 & IRD\\
2459892.9638331 & 5.67 & 4.04 & IRD\\
2459892.9686761 & 5.54 & 4.06 & IRD\\
2459893.9629992 & 8.26 & 3.60 & IRD\\
2459893.9706576 & 8.93 & 3.65 & IRD\\
2459894.9363576 & $-$1.51 & 3.86 & IRD\\
2459894.9423608 & $-$4.99 & 3.97 & IRD\\
2459896.9683733 & 0.64 & 3.50 & IRD\\
2459896.9790742 & 4.49 & 3.64 & IRD\\
2459897.8840425 & $-$4.65 & 4.15 & IRD\\
2459897.8877025 & 2.02 & 4.29 & IRD\\
2459907.8269676 & $-$1.94 & 3.61 & IRD\\
2459907.8318053 & $-$5.50 & 3.97 & IRD\\
2459908.8295062 & $-$6.44 & 4.11 & IRD\\
2459908.8329353 & $-$3.18 & 4.48 & IRD\\
2459950.7186219 & $-$4.38 & 4.14 & IRD\\
2459950.7234543 & $-$3.62 & 4.31 & IRD\\
\hline
2460123.5931062 & $-$0.54 & 1.69 & CARMENES\\
2460124.5932957 & 1.84 & 2.71 & CARMENES\\
2460125.6430641 & $-$0.46 & 1.27 & CARMENES\\
2460127.6440508 & $-$0.90 & 1.42 & CARMENES\\
2460133.6110069 & $-$1.38 & 2.40 & CARMENES\\
2460157.6029270 & 2.78 & 1.39 & CARMENES\\
2460158.5175814 & 0.89 & 1.96 & CARMENES\\
2460168.6194806 & $-$3.12 & 1.29 & CARMENES\\
2460170.6250763 & $-$3.13 & 1.30 & CARMENES\\
2460172.5947363 & $-$1.73 & 2.01 & CARMENES\\
2460176.6622728 & $-$3.82 & 1.55 & CARMENES\\
2460180.5947426 & 0.42 & 1.48 & CARMENES\\
2460184.5995223 & 5.11 & 2.78 & CARMENES\\
2460193.6313475 & 4.36 & 2.01 & CARMENES\\
2460195.5738182 & 5.26 & 1.86 & CARMENES\\
2460196.5756710 & 3.25 & 1.56 & CARMENES\\
2460198.6561354 & 0.68 & 1.72 & CARMENES\\
 \enddata
 \tablecomments{The BJDs, RVs, and RV errors were rounded from the ones originally provided by the pipelines. 
 }
\end{deluxetable}

\section{Odd-Even Transit Plot for the TESS Data}
\label{sec:odd_even}

The phase-folded TESS light curves of the odd- and even-numbered transits for the $P=12.76$\,day ephemeris are shown in Figure \ref{fig:odd_even}, which was truncated from the DV report produced by the SPOC pipeline \citep{2018PASP..130f4502T} from Sectors 40 -- 70.

\begin{figure*}[h]
   \centering
   \includegraphics[width=0.5\textwidth]{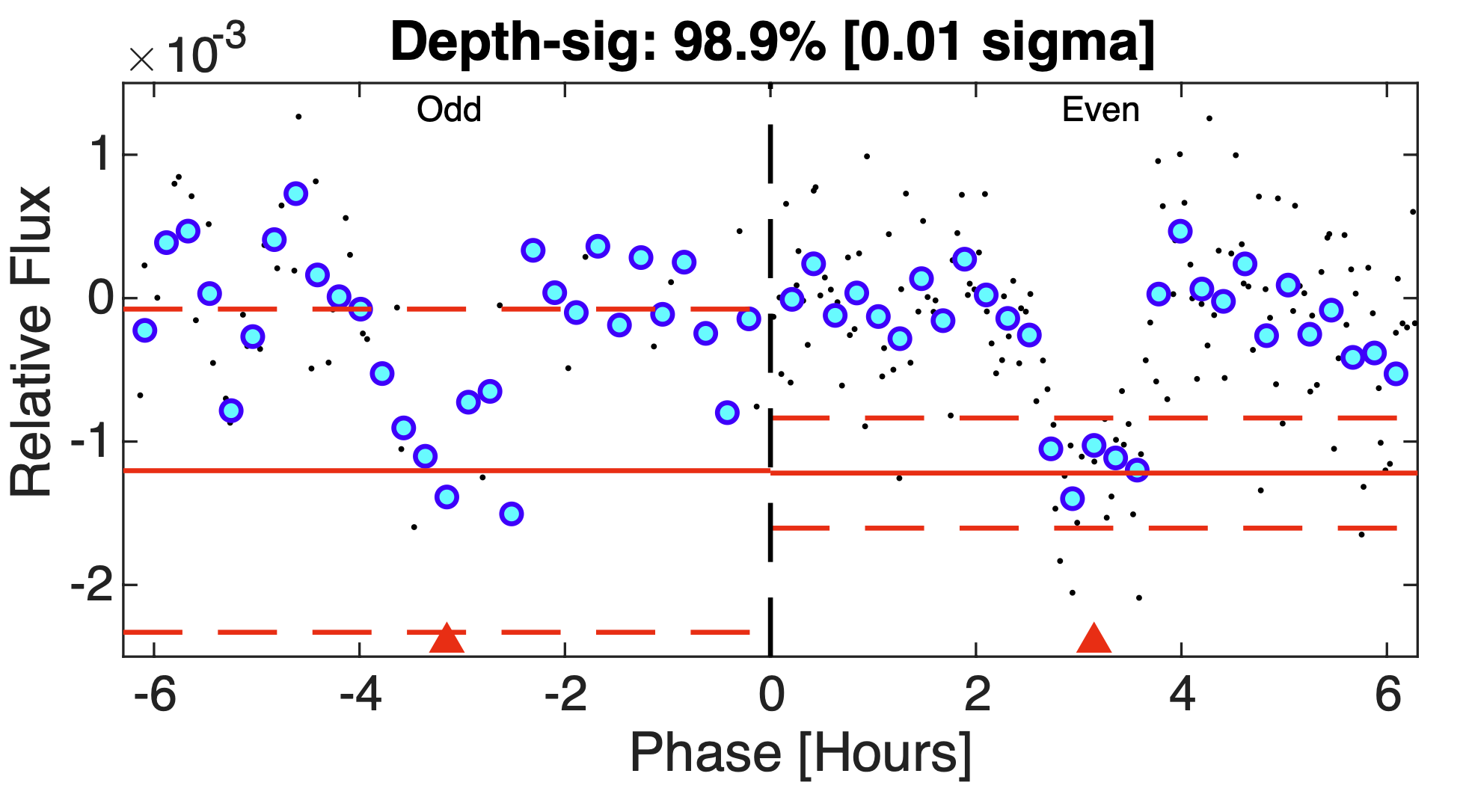}
   \caption{TESS light curves (black dots) of \target for the odd- (left) and even-numbered (right) transits. They were phase-folded for the $P=12.76$\,day ephemeris. Binned data points are shown as cyan-filled blue circles.
   The transit depths and 1$\sigma$ uncertainties were determined from independent transit model fits to each light curve and are shown as horizontal red solid and dashed lines. 
   The ``Depth-sig'' at the top of the plot is the statistical significance of the difference between the odd and even transit depths.
   The low significance (0.01$\sigma$) indicates there is no significant discrepancy in the measured depths between the two sets.
   The plots are reproduced from the DV report produced by the SPOC pipeline \citep{2018PASP..130f4502T} for Sectors 40 -- 70 (the full report is available from MAST). 
   }
   \label{fig:odd_even}
\end{figure*}

\section{Priors and Posteriors of the Light-curve Analyses}
\label{sec:lc_analysis}

Table~\ref{tab:lc_analysis} summarizes the priors and posteriors of the transit model parameters used for the analyses of light curves obtained with TESS, MuSCAT2, and MuSCAT3 (Sections \ref{sec:LC_TESS} and \ref{sec:LC_MuSCATs}).

\begin{deluxetable*}{lcccc}[h]
\tablecaption{Priors and posteriors of the parameters adopted in the transit modeling for the TESS and MuSCAT light curves\label{tab:lc_analysis}}
\tablehead{
        {Parameter} &  {Unit} & {Prior \tablenotemark{a}} & {Posterior (TESS)} & {Posterior (MuSCAT series)}  
}
\startdata
$b$ & & $\mathcal{U}$(0, 1) & $0.55 ^{+0.31}_{-0.37}$ & $0.60 ^{+0.21}_{-0.37}$\\
$\ln a_R$ & & $\mathcal{U}$(0, 5.3) & $4.23 ^{+0.18}_{-0.50}$ & $4.25 ^{+0.19}_{-0.30}$\\
$k$ $(=R_p/R_s)$ & & $\mathcal{U}$(0, 0.3) & $0.0320 ^{+0.0032}_{-0.0021}$ & $0.0326 ^{+0.0025}_{-0.0019}$\\
$T_{\rm c,0}$ & BJD & $\mathcal{U}$(0, $\infty$) & $2459497.1848 \pm 0.0036$ & $2459497.1842 ^{+0.0086}_{-0.0082}$\\
$P$ & days & $\mathcal{U}$(11.5, 14.0) & $12.76146 ^{+0.00009}_{-0.00011}$& $12.76145 \pm 0.00016$\\
$u_{\rm 1,TESS}$ &  & $\mathcal{N}$(0.261, 0.004) & $0.261 \pm 0.004$& ---\\
$u_{\rm 2, TESS}$ & & $\mathcal{N}$(0.308, 0.008) & $0.309 \pm 0.008$ & ---\\
$u_{1,r}$ & & $\mathcal{N}$(0.458, 0.011) & --- & $0.458 \pm 0.011$\\
$u_{2,r}$ & & $\mathcal{N}$(0.354, 0.017) & --- & $0.353 \pm 0.017$\\
$u_{1,i}$ & & $\mathcal{N}$(0.285, 0.006) & --- & $0.285 \pm 0.006$\\
$u_{2,i}$ & & $\mathcal{N}$(0.351, 0.011) & --- & $0.350 \pm 0.011$\\
$u_{1,z_s}$ & & $\mathcal{N}$(0.222, 0.005) & --- & $0.222 \pm 0.005$\\
$u_{2,z_s}$ & & $\mathcal{N}$(0.320, 0.010) & --- & $0.320 \pm 0.010$\\
\enddata
\tablenotetext{a}{$\mathcal{U}(a, b)$ denotes a uniform distribution between $a$ and $b$, and $\mathcal{N}(a, b)$ denotes a normal distribution with the mean of $a$ and variance of $b^2$.}
\end{deluxetable*}

\section{Posteriors of the joint-fit parameters}
\label{sec:posterior_of_joint-fit}

Figure~\ref{fig:corner_jointfit} shows a corner plot for the posterior probability distributions of selected transit+RV parameters calculated from the joint MCMC analysis in Section~\ref{sec:joint}. Figure~\ref{fig:TTVs} shows the individual midtransit times derived from the same analysis, where the best-fit linear ephemeris is subtracted.

\begin{figure*}[]
    \centering
    \includegraphics[width=14cm]{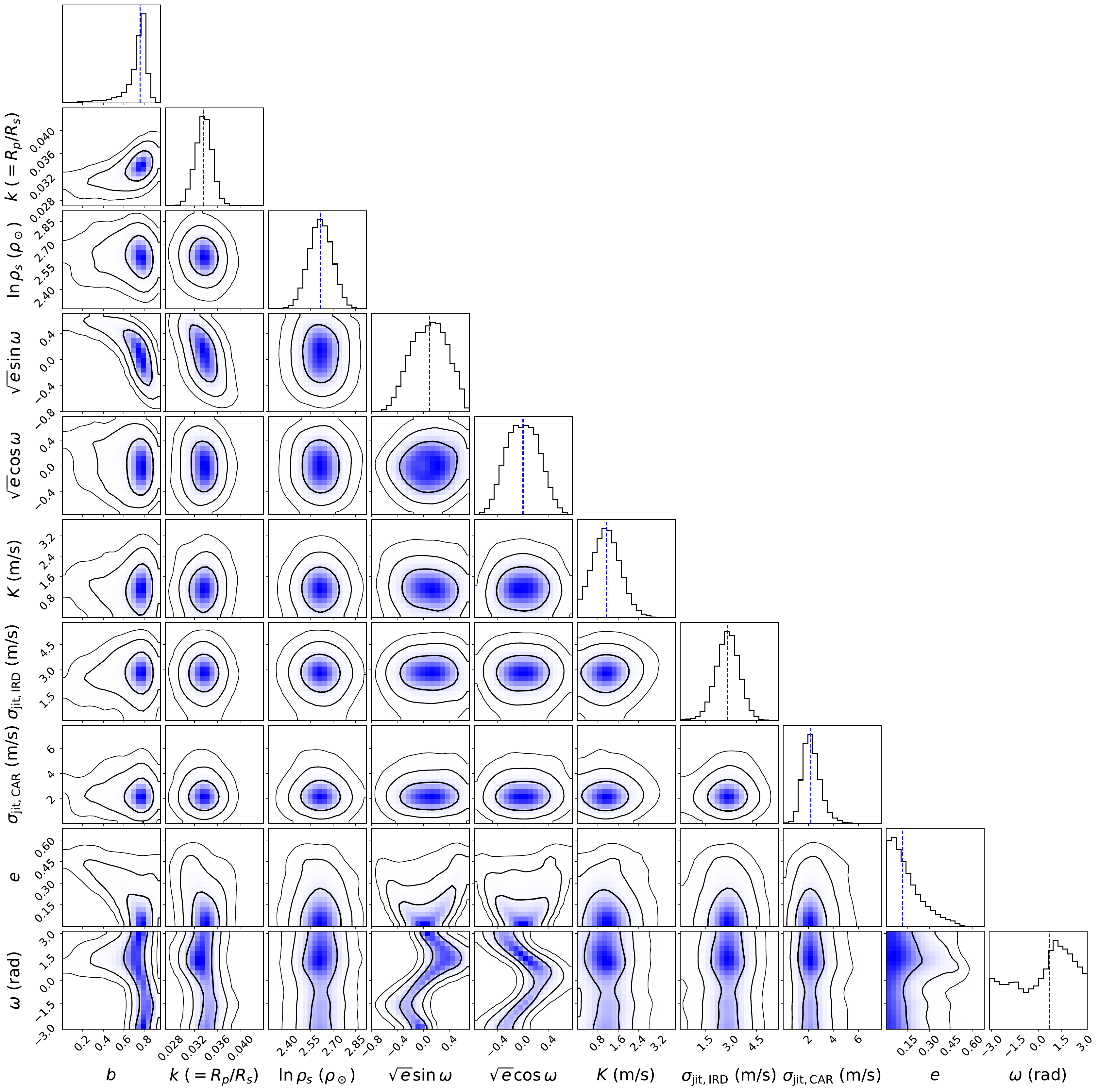}
    \caption{Corner plot for the posteriors of the joint-fit parameters calculated from the joint MCMC analysis.    
    The probability density is shown in blue, while 1$\sigma$, 2$\sigma$, and 3$\sigma$ confidence intervals are shown in the thickest, second-thickest, and thinnest black lines, respectively. The median value of each parameter is indicated by the vertical dashed line in the histograms.} 
    \label{fig:corner_jointfit}
\end{figure*}

\begin{figure}[]
    \centering
    \includegraphics[width=8cm]{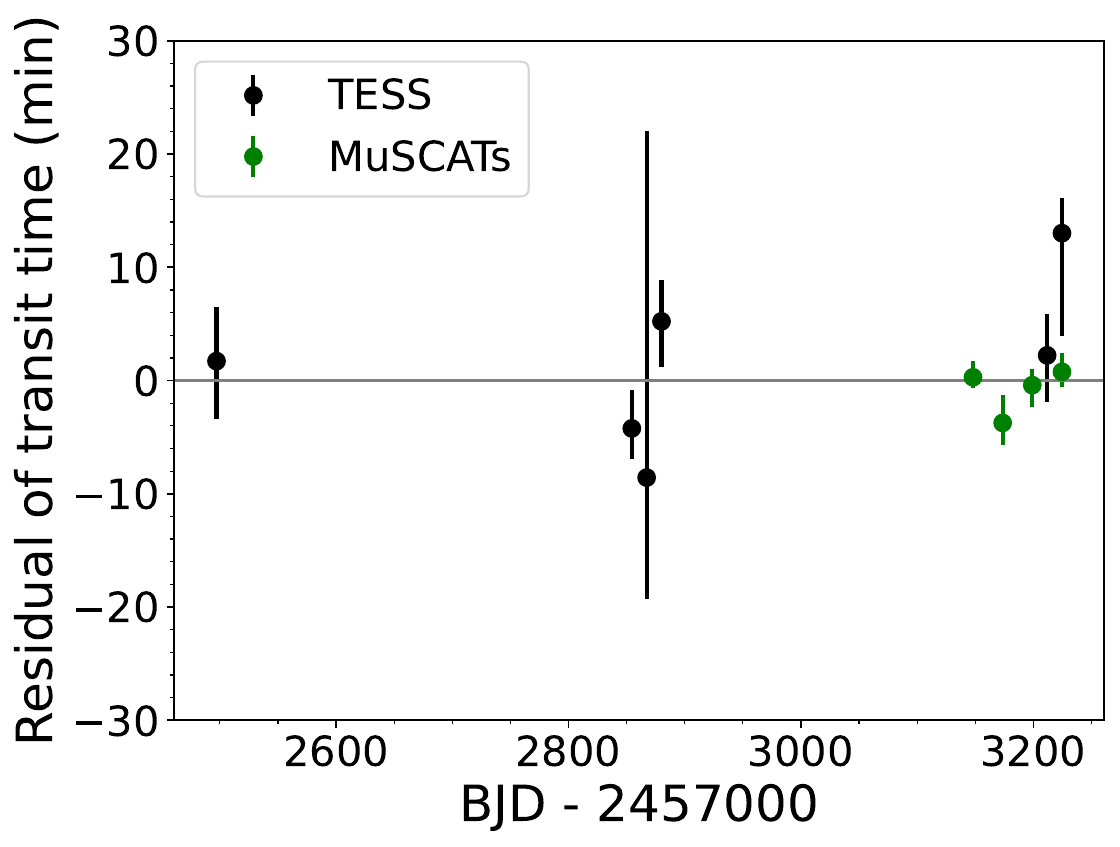}
    \caption{Midtransit times of \planet measured in the TESS (black) and MuSCAT series (green) data with the best-fit linear ephemeris subtracted, showing no significant TTVs. Note that the measurement with large error bars (the third data point from TESS) comes from a partial transit (transit epoch = 29).}
    \label{fig:TTVs}
\end{figure}


\end{document}